\documentclass[aps,prd,twocolumn,preprintnumbers,showpacs,nofootinbib,amssymb]{revtex4}
\usepackage{graphicx}
\usepackage{amsmath}
\usepackage{amssymb}
\usepackage{amsfonts}
\usepackage{bm}
\usepackage{color}

\def\be{\begin{equation}}
\def\ee{\end{equation}}
\def\bea{\begin{eqnarray}}
\def\eea{\end{eqnarray}}

\begin{document}

\title{Bose-Einstein Condensate general relativistic stars}
\author{Pierre-Henri Chavanis}
\email{chavanis@irsamc.ups-tlse.fr}
\affiliation{Laboratoire de Physique Th\'eorique (IRSAMC), CNRS and UPS,
Universit\'e de Toulouse, F-31062 Toulouse, France}
\author{Tiberiu Harko}
\email{harko@hkucc.hku.hk}
\affiliation{Department of Physics and
Center for Theoretical and Computational Physics, The University
of Hong Kong, Pok Fu Lam Road, Hong Kong, P. R. China}

\begin{abstract}
We analyze the possibility that due to their superfluid
properties some compact astrophysical objects may
contain a significant part of their matter in the form of a
Bose-Einstein condensate. To study the condensate we use the Gross-Pitaevskii
equation, with arbitrary non-linearity. By introducing the Madelung
representation of the wave function, we formulate the dynamics of the
system in terms of the continuity equation and of the hydrodynamic
Euler equations. The non-relativistic and Newtonian Bose-Einstein
gravitational condensate can be described as a gas, whose density and
pressure are related by a barotropic equation of state. In the case of
a condensate with quartic non-linearity, the equation of state is
polytropic with index one. In the framework of the Thomas-Fermi
approximation the structure of the Newtonian gravitational condensate
is described by the Lane-Emden equation, which can be exactly solved.
The case of the rotating condensate is also discussed. General
relativistic configurations with quartic non-linearity are studied
numerically with both non-relativistic and relativistic equations of
state, and the maximum mass of the stable configuration is
determined. Condensates with particle masses of the
order of two neutron masses (Cooper pair) and scattering length of the
order of $10-20$ fm have maximum masses of the order of $2M_{\odot }$,
maximum central density of the order of $0.1-0.3\times 10^{16}\, {\rm
g}/{\rm cm}^3$ and minimum radii in the range of $10-20$ km. In this way we
obtain a large class of stable astrophysical objects, whose basic
astrophysical parameters (mass and radius) sensitively depend on the
mass of the condensed particle, and on the scattering length. We also
propose that the recently observed neutron stars with masses in the
range of $2-2.4M_{\odot}$ are Bose-Einstein Condensate stars.
\end{abstract}


\maketitle

\section{Introduction}

At very low temperatures, particles in a dilute Bose gas can
occupy the same quantum ground state, forming a Bose-Einstein
(BEC) condensate, which appears as a sharp peak over a broader
distribution in both coordinates and momentum space. The
possibility to obtain quantum degenerate gases by a combination of
laser and evaporative cooling techniques has opened several new
lines of research, at the border of atomic, statistical and
condensed matter physics (for recent reviews see
\cite{Da99, Pet}).

To say that so many particles are in the same quantum state is equivalent in
saying that these particles display the state coherence. That is, BEC is a
particular case of coherence phenomena, related to the arising state
coherence. As the gas is cooled, the condensation of a large fraction of the
particle in a gas occurs via a phase transition, taking place when the
wavelengths of individual particles overlap and behave identically. For the
transition to take place, particles have to be strongly correlated with each
other \cite{Da99, Pet}.

For an ensemble of particles in thermodynamic equilibrium at temperature $T$, the thermal energy of a particle is given by $k_{B}T$, where $k_{B}$ is
Boltzmann's constant. For a particle of mass $m$, the thermal wavelength is $\lambda _{T}=\sqrt{2\pi \hbar ^{2}/mk_{B}T}$. Particles become
correlated with each other when their wavelengths overlap, that
is, the thermal wavelength is greater than the mean
inter-particles distance $l$, $\lambda
_{T}>l$. The average particle number $n$ for $N$ particles in a volume $V$, $n=N/V$, is related to the distance $l$ through the relation
$nl^{3}=1$. Hence the condition $\lambda _{T}>l$ can be rewritten
as $n\lambda _{T}^{3}>1 $, which yields the inequality
\cite{LaLi80}
\begin{equation}
T<\frac{2\pi \hbar ^{2}}{mk_{B}}n^{2/3}.
\end{equation}

Hence a coherent state may develop if the particle density is high
enough or the temperature is sufficiently low. An accurate description
of the BEC for an ideal gas is based on the Bose-Einstein distribution
$f\left( p\right) =\left\{ \exp \left[ \left( \varepsilon _{p}-\mu
\right) /k_{B}T\right] -1\right\} ^{-1}$, for particles with momentum
$p$, energy $\varepsilon _{p}=p^{2}/2m$ and chemical potential $\mu
$. In the thermodynamic limit $N\rightarrow \infty $, $V\rightarrow
\infty $, $N/V\rightarrow $constant, the fraction of particles
condensing to the state with $p=0$ below the condensation temperature
$T_{c}$ is $n_{0}=1-\left( T/T_{c}\right) ^{3/2}$, while $n_{0}=0$
above the condensation temperature. The condensation temperature is
$T_{c}=2\pi \hbar ^{2}n^{2/3}/mk_{B}\zeta ^{2/3}$, where $\zeta
=2.612$ \citep{LaLi80}. The dynamical process of Bose-Einstein
condensation in the canonical ensemble (fixed temperature $T$) has
been studied in \cite{sopik}.

A non-ideal, weakly interacting Bose gas also displays Bose-Einstein
condensation, though particles interactions deplete the condensate, so that
at zero temperature the condensate fraction is smaller than unity, $n_{0}<1$. A system is called weakly interacting if the characteristic
interaction radius $r_{int}$ is much smaller than the mean
inter-particles distance $l$,
$r_{int}\ll l$. This inequality can be rewritten equivalently as $nr_{int}^{3}\ll 1$. If this condition holds, the system is called dilute \cite{Pet}.

Superfluid liquids, like $^{4}$He, are far from being dilute.
Nevertheless, one believes that the phenomenon of superfluidity is
related with BEC. The experimental observations and the
theoretical calculations estimate the
condensate fraction for superfluid helium at $T=0$ to be $n_{0}\approx 0.10$. A strongly correlated pair of fermions can be treated
approximately like a boson. This is why the arising superfluidity
in $^{3}$He can be interpreted as the condensation of coupled
fermions. Similarly, superconductivity may be described as the
condensation of the Cooper pairs that are formed by the electrons
or the holes \citep{Ch05}.

An ideal system for the experimental observation of the BEC
condensation is a dilute atomic Bose gas confined in a trap and
cooled to very low temperatures. BEC were first observed in 1995
in dilute alkali gases such as vapors of rubidium and sodium. In
these experiments, atoms were confined in magnetic traps,
evaporatively cooled down to a fraction of a microkelvin, left to
expand by switching off the magnetic trap, and subsequently imaged
with optical methods. A sharp peak in the velocity distribution
was observed below a critical temperature, indicating that
condensation has occurred, with the alkali atoms condensed in the
same ground state. Under the typical confining conditions of
experimental settings, BECs are inhomogeneous, and hence
condensates arise as a narrow peak not only in the momentum space
but also in the coordinate space \cite{An95,Br95,Da95}.

If considering only two-body, mean field interactions, a dilute
Bose-Einstein gas near zero temperature can be modelled using a cubic
non-linear Schr\"odinger equation with an external potential, which is known
as the Gross-Pitaevskii equation \cite{Pet}.

The possibility of the Bose-Einstein condensation has also been
considered in nuclear and quark matter, in the framework of the
analysis of the BCS-BEC crossover. At ultra-high density, matter is expected to form
a degenerate Fermi gas of quarks in which the Cooper pairs of
quarks condensate near the Fermi surface (color superconductor).
If the attractive interaction is strong enough, at some critical
temperature the fermions may condense into the bosonic zero mode,
forming a Bose-Einstein quark condensate
\cite{Bal95}. The basic concept of the BCS-BEC
crossover is as follows: As long as the attractive interaction
between fermions is weak, the system exhibits the superfluidity
characterized by the energy gap in the BCS mechanism. On the other
hand, if the attractive interaction is strong enough, the fermions
first form bound molecules (bosons). Then they start to condense
into the bosonic zero mode at some critical temperature. These two
situations are smoothly connected without a phase transition
\citep{NiAb05}.

One of the most striking features of the crossover is that the
critical temperature in the BEC region is independent of the
coupling for the attraction between fermions. This is because the
increase of the coupling only affects the internal structure of
the bosons, while the critical temperature is determined by the
boson's kinetic energy. Thus, the critical temperature reaches a
ceiling for the large coupling as long as the binding effect on
the boson mass can be neglected. Even in the nuclear matter where
the interaction is relatively strong, the binding energy of the
deuteron is much smaller than the nucleon mass. This fact allows
us to work within a non-relativistic framework to describe such a
crossover \citep{NiAb05}. However, in relativistic systems
where the binding energy cannot be neglected, there could be two
crossovers in the relativistic fluids: one is the ordinary BCS-BEC
crossover, where the critical temperature in the BEC region would
not plateau because of the relativistic effect, and the second is
the crossover from the BEC state to a relativistic state, the
so-called relativistic BEC (RBEC), where the critical temperature
increases to the order of the Fermi energy \citep{NiAb05}.

In isospin symmetric nuclear matter,
neutron-proton ($np$) pairing undergoes a smooth transition
leading from an assembly of $np$ Cooper pairs at higher densities
to a gas of Bose-condensed deuterons as the nucleon density is
reduced to an extremely low value. This transition may be relevant
to supernova matter or for the crust of neutron stars
\cite{SeCl06}. A mixture of
interacting neutral and charged Bose condensates, which is
supposed to be realized in the interior of neutron stars in the
form of a coexistent neutron superfluid and protonic
superconductor, was considered in \cite{Ba04}.

The possibility of the existence of some Bose condensates in
neutron stars was considered for a long time (see Glendenning \cite{Gl00} for
a detailed discussion). The condensation of negatively charged
mesons in neutron star matter is favored because such mesons would
replace electrons with very high Fermi momenta. The in-medium
properties of the $K^{-}$ mesons may be such that they could
condense in neutron matter as well. Bose-Einstein condensates of
kaons/anti-kaons in compact objects were discussed recently
\cite{BaBa03,Ban04}. Pion as well as kaon condensates would have
two important effects on neutron stars. Firstly, condensates
soften the equation of state above the critical density for onset
of condensation, which reduces the maximal possible neutron star
mass. At the same time, however, the central stellar density
increases, due to the softening. Secondly, meson condensates would
lead to neutrino luminosities which are considerably enhanced over
those of normal neutron star matter. This would speed up neutron
star cooling considerably \citep{Gl00}. Another particle which may
form a condensate is the H-dibaryon, a doubly strange six quark
composite with spin and isospin zero, and baryon number two. In
neutron star matter, which may contain a significant fraction of
$\Lambda $ hyperons, the $\Lambda $'s could combine to form
H-dibaryons. H-matter condensates may thus exist at the center of
neutron stars \citep{Gl00}. Neutrino superfluidity, as suggested
by Kapusta \cite{Ka04}, may also lead to Bose-Einstein condensation
\citep{Ab06}.

Zero spin bosons, described by real or complex scalar fields, are the
simplest particles which can be considered in the framework of quantum
field theory and general relativity.  Real scalar fields have
equilibrium configuration that were discovered by Seidel and Suen
\cite{SeSu91} and are called oscillatons. They are globally regular
but are fully time dependent. As for their stability, they seem to be
quite robust as far as numerical evolution is concerned
\cite{Al04}. The objects which can be formed by scalar fields have
been investigated in detail by using mainly numerical tools
\cite{Al04-1}.  Complex scalar fields can form stable equilibrium
configurations called boson stars \cite{bosonstars,colpi}, that are globally
regular and whose energy density is time independent. The possibility
that dark matter is in the form of a scalar field
\cite{scalarfield,goodman,arbey} or a Bose-Einstein condensate  \cite{dark} has
also been investigated extensively.

Therefore the physical results presented above show that the
possibility of the existence of a Bose-Einstein condensate inside
compact astrophysical objects or the existence of stars formed
entirely from a Bose-Einstein condensate cannot be excluded {\it a
priori}. Such a possibility has been in fact suggested recently.
Wang \cite{Wa01} used the Gross-Pitaevskii equation, together with the
associated energy functional and the Thomas-Fermi approximation, to
study a cold star composed of a dilute Bose-Einstein condensate.
For a static star, the exact solution for the density distribution
was obtained. A number of perturbative solutions for the case of a
slowly rotating star have also been derived. The effect of a
scalar dark matter background on the equilibrium of degenerate
stars was studied by Grifols \cite{Gr06}, with a particular focus on
white dwarfs, and the changes induced in their masses and radii.

A detailed analytical and numerical analysis of the Newtonian
Bose-Einstein condensate systems was performed recently in
\cite{Chav1} and \cite{Chav2}, respectively. In \cite{Chav1} an
approximate analytical expression of the mass-radius relation of a
Newtonian self-gravitating Bose-Einstein condensate with short-range
interactions, described by the Gross-Pitaevskii-Poisson system, was
obtained. For repulsive short-range interactions (positive scattering
lengths), configurations of arbitrary mass do exist, but their radius
is always larger than a minimum value. For attractive short-range
interactions (negative scattering lengths), equilibrium configurations
only exist below a maximum mass. The equation of hydrostatic
equilibrium describing the balance between the gravitational
attraction and the pressure due to quantum effects and short-range
interactions (scattering) was numerically solved in \cite{Chav2}.

It is the purpose of the present paper to develop a general and
systematic formalism for the study of gravitationally bounded
Bose-Einstein condensates, in both Newtonian and general relativistic
situations. Our approach is independent of the nature of the
condensate. As a starting point we generalize the Gross-Pitaevskii
equation by allowing an arbitrary form of the non-linearity. To obtain
a transparent description of the physical properties of the BECs we
introduce the hydrodynamical representation of the wave function,
which allows the formulation of the dynamics of the condensate in
terms of the continuity and hydrodynamic Euler equations. Hence the
Bose-Einstein gravitational condensate can be described as a gas whose
density and pressure are related by a barotropic equation of state. In
the case of a condensate with quartic non-linearity, the equation of
state of the condensate is given by a polytropic equation of state
with polytropic index $n=1$. In the framework of the Thomas-Fermi
approximation, with the quantum potential neglected, the structure of
the gravitational BEC is described by the Lane-Emden equation, which
can be solved analytically. Hence the mass and the radius of the
condensate can be easily obtained. The case of the rotating Newtonian
condensate is also discussed, by using the generalized Lane-Emden
equation.

By using the equation of state corresponding to the Bose-Einstein
condensates with quartic non-linearity we consider the general
relativistic properties of condensate stars, by numerically
integrating the structure equations (the mass continuity and the
Tolman-Oppenheimer-Volkoff equation) for a static configuration. In
our general relativistic study we consider the cases of condensates
described by both non-relativistic and relativistic equations of
state, respectively. The maximum mass and the corresponding radius are
obtained numerically. Bose-Einstein condensate stars
with particle masses of the order of two neutron masses (Cooper pair)
and scattering length of the order of $10-20$ fm have maximum masses
of the order of $2M_{\odot}$, maximum central density of the order of
$0.1-0.3\times 10^{16}\, {\rm g}/{\rm cm}^3$ and minimum radii in the range of
$10-20$ km.

The present paper is organized as follows. The Gross-Pitaevskii
equation is written down in Section \ref{sect2}. The hydrodynamical
representation for the study of the gravitationally bounded BECs is
introduced in Section \ref{sect3}. The static and slowly rotating
Newtonian condensates are analyzed in Section \ref{sect4}. The maximum
mass of Newtonian condensate stars is discussed in Section
\ref{sect5}. The properties of the general relativistic static
condensates with quartic non-linearity are studied in Section
\ref{sect6} for both non-relativistic and relativistic equations of state,
respectively. The astrophysical implications of our results are
considered in Section \ref{sect7}. We discuss and conclude our results
in Section \ref{sect8}.

\section{The Gross-Pitaevskii equation for the Bose-Einstein condensate stars}\label{sect2}

In a quantum system of $N$ interacting condensed bosons most of the bosons
lie in the same single-particle quantum state. The many-body Hamiltonian
describing the interacting bosons confined by an external potential $V_{ext}$
is given, in the second quantization, by
\begin{eqnarray}
&&\hat{H}=\int d\vec{r}\hat{\Psi}^{+}\left( \vec{r}\right) \left[ -\frac{\hbar ^{2}}{2m}\nabla ^{2}+V_{rot}\left( \vec{r}\right) +V_{ext}\left( \vec{r}\right) \right] \hat{\Psi}\left( \vec{r}\right) \nonumber \\
&&+\frac{1}{2}\int d\vec{r}d\vec{r}^{\prime }\hat{\Psi}^{+}\left( \vec{r}\right) \hat{\Psi}^{+}\left( \vec{r}^{\prime }\right) V\left( \vec{r}-\vec{r}^{\prime }\right) \hat{\Psi}\left( \vec{r}\right) \hat{\Psi}\left( \vec{r}^{\prime }\right) ,  \label{ham}
\end{eqnarray}
where $\hat{\Psi}\left( \vec{r}\right) $ and $\hat{\Psi}^{+}\left( \vec{r}\right) $ are the boson field operators that annihilate and create a
particle at the position $\vec{r}$, respectively, and $V\left( \vec{r}-\vec{r}^{\prime }\right) $ is the two-body interatomic potential \cite{Pet}. $V_{rot}\left( \vec{r}\right) $ is the potential associated to the rotation
of the condensate, and is given by
\begin{equation}
V_{rot}\left( \vec{r}\right) =f_{rot}\left( t\right) \frac{m\omega ^{2}}{2}{r}^{2},
\end{equation}
where $\omega $ is the angular velocity of the condensate and $f_{rot}\left(
t\right) $ a function which takes into account the possible time variation
of the rotation potential. For a system consisting of a large number of
particles, the calculation of the ground state of the system with the direct
use of Eq.~(\ref{ham}) is impracticable, due to the high computational cost.

Therefore the use of some approximate methods can lead to a significant
simplification of the formalism. One such approach is the mean field
description of the condensate, which is based on the idea of separating out
the condensate contribution to the bosonic field operator. For a uniform gas
in a volume $V$, BEC occurs in the single particle state $\Psi _{0}=\sqrt{N/V}
$, having zero momentum. The field operator can then be decomposed  in the
form $\hat{\Psi}\left( \vec{r}\right) =\sqrt{N/V}+\hat{\Psi}^{\prime }\left(\vec{r}\right) $. By treating the operator $\hat{\Psi}^{\prime }\left( \vec{r}\right) $ as a small perturbation, one can develop the first order theory
for the excitations of the interacting Bose gases \cite{Da99,Pet}.

In the general case of a non-uniform and time-dependent configuration, the
field operator in the Heisenberg representation is given by
\begin{equation}
\hat{\Psi}\left( \vec{r},t\right) =\psi \left( \vec{r},t\right) +\hat{\Psi}^{\prime }\left( \vec{r},t\right) ,
\end{equation}
where $\psi \left( \vec{r},t\right) $, called the condensate wave
function, is the expectation value of the field operator, $\psi \left( \vec{r},t\right) =\langle \hat{\Psi}\left( \vec{r},t\right) \rangle $.
It is a classical field and its absolute value fixes the number density of
the condensate through $n \left( \vec{r},t\right) =\left| \psi \left(
\vec{r},t\right) \right| ^{2}$.  The normalization condition is $N=\int n
\left( \vec{r},t\right) d\vec{r}$, where $N$ is the total number of
particles in the star.

The equation of motion for the condensate wave function is given by the
Heisenberg equation corresponding to the many-body Hamiltonian given by Eq.~(\ref{ham}),
\begin{eqnarray}
&&i\hbar \frac{\partial }{\partial t}\hat{\Psi}\left( \vec{r},t\right) =\left[ \hat{\Psi},\hat{H}\right] =  \nonumber \\
&&\left[ -\frac{\hbar ^{2}}{2m}\nabla ^{2}+V_{rot}\left( \vec{r}\right)
+V_{ext}\left( \vec{r}\right) \right. +\nonumber\\
&&\left.\int d\vec{r}^{\prime }\hat{\Psi}^{+}\left(
\vec{r}^{\prime },t\right) V\left( \vec{r}^{\prime }-\vec{r}\right) \hat{\Psi}\left( \vec{r}^{\prime },t\right) \right] \hat{\Psi}\left( \vec{r},t\right)
.  \label{gp}
\end{eqnarray}

Replacing $\hat{\Psi}\left( \vec{r},t\right) $ by the condensate wave
function $\psi $ gives the zeroth-order approximation to the Heisenberg
equation. In the integral containing the particle-particle interaction $V\left( \vec{r}^{\prime }-\vec{r}\right) $ this replacement is in general a
poor approximation for short distances. However, in a dilute and cold gas,
only binary collisions at low energy are relevant and these collisions are
characterized by a single parameter, the $s$-wave scattering length,
independently of the details of the two-body potential. Therefore, one can
replace $V\left( \vec{r}^{\prime }-\vec{r}\right) $ by an effective
interaction $V\left( \vec{r}^{\prime }-\vec{r}\right) =\lambda \delta \left(
\vec{r}^{\prime }-\vec{r}\right) $, where the coupling constant $\lambda $
is related to the scattering length $a$ through $\lambda =4\pi \hbar ^{2}a/m$. Hence, we assume that in a medium composed of scalar particles
with non-zero mass, the range of Van der Waals-type scalar
mediated interactions among nucleons becomes infinite, when the
medium makes a transition to a Bose-Einstein condensed phase.

With the use of the effective potential the integral in the bracket of Eq. (\ref{gp}) gives $\lambda \left| \psi \left( \vec{r},t\right) \right| ^{2}$,
and the resulting equation is the Schr\"odinger equation with a quartic
nonlinear term \cite{Da99, Pet}. However, in order to obtain a more general
description of the Bose-Einstein condensate stars, we shall assume an
arbitrary non-linear term $g^{\prime }( \left| \psi ( \vec{r},t)
\right| ^{2}) $ \cite{Ko00}. As was pointed out in \cite{Ko00}, the
Gross-Pitaevskii approximation is a long-wavelength theory widely used to
describe a variety of properties of dilute Bose condensates, but for
short-ranged repulsive interactions this theory fails in low dimensions, and
some essential modifications of the theory are necessary.

From a physical point of view these modifications can be understood as
follows \cite{Ko00}. The inter-particle interaction can be written as $V\left( \vec{r}\right) =u_{0}\delta _{a}^{d}\left( \vec{r}\right) $, where $u_{0}$ is the amplitude of the inter-particle repulsion and
$\delta _{a}^{d}\left( \vec{r}\right) $ denotes any well localized
$d$-dimensional function that transforms into the mathematical
Dirac delta distribution when the range of interactions $a\rightarrow
0$. Assume that the inter-particle interaction is so strong that
each particle is localized within a cage formed by its neighbors.
In the dilute limit $nl^{d}\ll 1$ the size of this cage can be
estimated as $R\sim n^{-1/d}$ and the ground state energy per
particle follows from the uncertainty principle as $\hbar
^{2}/mR^{2}\sim \hbar ^{2}n^{2/d}/m$. The ground state energy
which would go into the energy functional is given by $\hbar
^{2}n^{\left( 2+d\right) /d}/m$. The strong
interaction assumption is valid if the interaction energy per particle $u_{0}/R^{d}$ is much bigger than the ground state energy per particle, i.e. $u_{0}/R^{d}\gg$ $\hbar ^{2}/mR^{2}$. The condition for the strong
coupling limit can be written as $\hbar ^{2}n^{\left( 2-d\right)
/d}/mu_{0}\ll 1$. As space dimensionality decreases, it becomes
increasingly harder for the repulsive particles to avoid
collisions. Thus, in general, the quartic non-linearity in the
energy functional should be replaced by $\left| \psi \right|
^{2\left( 2+d\right) /d}$. The non-linearity may also be
logarithmic in $\left| \psi \right| ^{2}$ \cite{Ko00}. For $d=1$
we have a $\left| \psi \right| ^{6}$ interaction.

Therefore the generalized Gross-Pitaevskii equation describing a
gravitationally trapped Bose-Einstein condensate is given by
\begin{eqnarray}  \label{gen}
i \hbar \frac{\partial }{\partial t}\psi \left(
\vec{r},t\right) &=&\left[ -\frac{\hbar ^{2}}{2m}\nabla ^{2}+V_{rot}\left(
\vec{r}\right) +V_{ext}\left( \vec{r}\right) +\right. \nonumber\\
&&  g^{\prime}\left( \left| \psi
\left( \vec{r},t\right) \right| ^{2}\right)  \Big] \psi \left( \vec{r},t\right),
\end{eqnarray}
where we denoted $g^{\prime}=dg/dn $.

As for $V_{ext}\left( \vec{r}\right) $, we assume that it is the
gravitational potential, $V_{ext}=m \Phi$, and it satisfies the Poisson
equation
\begin{equation}
\nabla ^{2}\Phi=4\pi G\rho ,  \label{pot}
\end{equation}
where
\begin{equation}
\rho =mn=m\left| \psi
\left( \vec{r},t\right) \right| ^{2},
\end{equation}
is the mass density inside the Bose-Einstein condensate star.

\section{The hydrodynamical representation of the Bose-Einstein
gravitational condensate}\label{sect3}

The physical properties of a Bose-Einstein condensate described by the
generalized Gross-Pitaevskii equation given by Eq.~(\ref{gen}) can be
understood much easily by using the so-called Madelung representation of the
wave function \cite{Da99,Pet}, which consists in writing $\psi $ in the form
\begin{equation}
\psi \left( \vec{r},t\right) =\sqrt{n \left( \vec{r},t\right) }\exp \left[
\frac{i}{\hbar }S\left( \vec{r},t\right) \right] ,
\end{equation}
where the function $S\left( \vec{r},t\right) $ has the dimension of an
action. By substituting the above expression of the wave function into Eq.~(\ref{gen}) it decouples into a system of two differential equations for the
real functions $n $ and $\vec{v}$, given by
\begin{equation}
 \frac{\partial n }{\partial t}+\nabla \cdot \left( n \vec{v}\right) =0,
\end{equation}
\begin{equation}
 m\frac{\partial \vec{v}}{\partial t}+\nabla \left( \frac{m\vec{v}^{2}}{2}+V_{Q}+V_{rot}+V_{ext}+g^{\prime
}\right) =0,  \label{euler}
\end{equation}
where we have introduced the quantum potential
\begin{equation}
V_{Q}=-\frac{\hbar ^{2}}{2m}\frac{\nabla ^{2}\sqrt{n }}{\sqrt{n }},
\end{equation}
and the velocity of the quantum fluid
\begin{equation}
\vec{v}=\frac{\nabla S}{m},
\end{equation}
respectively. From its definition it follows that the velocity field is
irrotational, satisfying the condition $\nabla \times \vec{v}=\vec{0}$.

The quantum potential $V_{Q}$ has the property \cite{Da99}
\begin{equation}
n \nabla _{i}V_{Q}=\nabla _{j}\left( -\frac{\hbar ^{2}}{4m}n \nabla
_{i}\nabla _{j}\ln n \right) =\nabla _{j}\sigma _{ij}^{Q},
\end{equation}
where $\sigma _{ij}^{Q}=-(\hbar ^{2}n/4m) \nabla _{i}\nabla _{j}\ln n$ is the quantum stress tensor, which has the dimension of a pressure
and is an intrinsically anisotropic quantum contribution to the equations of
motion.

By taking into account that the flow
is irrotational, the equations of motion of the gravitational
ideal Bose-Einstein condensate take the form of the equation of
continuity and of the hydrodynamic Euler equation,
\begin{equation}
\frac{\partial \rho }{\partial t}+\nabla \cdot \left( \rho \vec{v}\right) =0,
\end{equation}
\begin{eqnarray}
\rho \left[ \frac{\partial \vec{v}}{\partial t}+\left( \vec{v}\cdot
\nabla \right) \vec{v}\right] =&-&\nabla P\left( \frac{\rho }{m}\right)
-\rho \nabla \left( \frac{V_{rot}}{m}\right)\nonumber\\
&-&\rho \nabla \left(
\frac{V_{ext}}{m}\right) -\nabla \cdot \sigma ^{Q},
\end{eqnarray}
where we have denoted
\begin{equation}
P\left( \frac{\rho }{m}\right) =g^{\prime }\left( \frac{\rho }{m}\right) \frac{\rho }{m}-g\left( \frac{\rho }{m}\right) .
\label{state}
\end{equation}
Therefore the Bose-Einstein gravitational condensate can be described as a
gas whose density and pressure are related by a barotropic equation of state \cite{Da99}. The explicit form of this equation depends on the form of the
non-linearity term $g$.

For a static ideal condensate, $\vec{v}\equiv \vec{0}$. In this case, from Eq.~(\ref
{euler}), we obtain
\begin{equation}
V_{Q}+V_{rot}+V_{ext}+g^{\prime }={\rm constant} \label{cons}
\end{equation}
Applying the operator $\nabla ^{2}$ to both sides of Eq.~(\ref{cons})
gives
\begin{equation}\label{nabla}
\nabla ^{2}\left( V_{Q}+V_{rot}+g^{\prime }\right) +\nabla ^{2}V_{ext}=0.
\end{equation}
In the case of a condensate with a non-linearity of the form $g(n) =k_0 m^2 n^{2}/2$, where  $k_0$ is a  constant,  and in the presence of a confining
gravitational field $V_{ext}=m\Phi$, it follows that the generalized potential $V_{gen}=-V_{Q}-V_{rot}-k_0 m \rho $ satisfies the Poisson equation,
\begin{equation}
\frac{1}{m}\nabla ^{2}V_{gen}=4\pi G\rho .
\end{equation}
If the quantum potential can be neglected, then from Eq.
(\ref{nabla}), by using the relations $\nabla ^{2}V_{rot}=m\omega
^2$ and $\nabla ^{2}g^{\prime }=k_0 m \nabla ^{2}\rho $, it follows
that the mass density of the condensate is described by a
Helmholtz type equation, given by
\begin{equation}
\nabla ^{2}\rho +\frac{4\pi G}{k_0}\rho +\frac{\omega ^{2}}{k_0}=0.
\end{equation}

\section{Static and slowly rotating Newtonian Bose-Einstein condensate stars}\label{sect4}

When the number of particles in the gravitationally bounded
Bose-Einstein condensate becomes large enough, the quantum
pressure term makes a significant contribution only near the
boundary of the condensate. Hence it is much smaller than the
non-linear interaction term. Thus the quantum stress term in the
equation of motion of the condensate can be neglected. This is the
Thomas-Fermi approximation, which has been extensively used for
the study of the Bose-Einstein condensates \citep{Da99}. As the
number of particles in the condensate becomes infinite, the
Thomas-Fermi approximation becomes exact \citep{Wa01,Chav1,Chav2}. This
approximation also corresponds to the classical limit of the
theory (it corresponds to neglecting all terms with powers of
$\hbar$) or as the regime of strong repulsive interactions among
particles. From a mathematical point of view the Thomas-Fermi
approximation corresponds to neglecting all terms containing $\nabla{\rho}$ and $\nabla{S}$ in the equation of motion.

\subsection{Static Bose-Einstein Condensate stars}

In the case of a static Bose-Einstein condensate, all physical
quantities are independent of time. Moreover, in the first
approximation we also neglect the rotation of the star, taking
$V_{rot}=0$. Therefore the equations describing the static
Bose-Einstein condensate  in a gravitational field with potential $V_{ext}=m\Phi$ take
the form
\begin{equation}
\nabla P=-\rho \nabla \Phi,
\end{equation}
\begin{equation}
\nabla ^{2}\Phi=4\pi G\rho .
\end{equation}

These equations must be integrated together with the equation of state $P=P\left( \rho \right) $, which follows from Eq. (\ref{state}), and some
appropriately chosen boundary conditions. By assuming that the non-linearity
in the Gross-Pitaevskii equation is of the form
\begin{equation}
g\left(n \right) =\alpha n ^{\gamma },
\end{equation}
where $\alpha $ and $\gamma $ are positive constants, it follows that the equation of
state of the gravitational Bose-Einstein condensate is the polytropic
equation of state,
\begin{equation}
P\left( \rho \right) =\alpha \left( \gamma -1\right) n^{\gamma
}=K\rho ^{\gamma },
\end{equation}
where we denoted $K=\alpha \left( \gamma -1\right)/m^{\gamma}$.

By representing $\gamma $ in the form $\gamma =1+1/n$, where $n$ is the
polytropic index, it follows that the structure of the static Bose-Einstein
condensate star is described by the Lane-Emden equation,
\begin{equation}
\label{laneemden}
\frac{1}{\xi ^{2}}\frac{d}{d\xi }\left (\xi ^{2}\frac{d\theta }{d\xi }\right )=-\theta ^{n},
\end{equation}
where $\theta $ is a dimensionless variable defined via $\rho =\rho
_{c}\theta ^{n}$, $\xi $ is a dimensionless coordinate introduced via
the transformation $r=[ (n+1)K\rho _{c}^{1/n-1}/4\pi G]
^{1/2}\xi $ and $\rho _{c}$ is the central density of the condensate \cite{Ch57}.

Hence the mass and the radius of the condensate are given by
\begin{equation}
\label{radiusn}
R=\left[ \frac{(n+1)K}{4\pi G}\right]^{1/2}\rho
_{c}^{(1-n)/2n}\xi _{1},
\end{equation}
and
\begin{equation}
\label{massn}
M=4\pi \left[ \frac{(n+1)K}{4\pi G}\right] ^{3/2}\rho
_{c}^{(3-n)/2n}\xi _{1}^{2}\left| \theta ^{\prime }\left( \xi _{1}\right)
\right| ,
\end{equation}
respectively, where $\xi _{1}$ defines the zero- pressure and zero-density
surface of the condensate: $\theta \left( \xi _{1}\right) =0$ \citep{Ch57}.

In the standard approach to the Bose-Einstein condensates, the non-linearity
term $g$ is given by
\begin{equation}
g\left(n \right) =\frac{u_{0}}{2}\left| \psi \right| ^{4}=\frac{u_{0}}{2}n ^{2},
\end{equation}
where $u_{0}=4\pi \hbar ^{2}a/m$ \cite{Da99}. The corresponding equation of
state of the condensate is
\begin{equation}\label{eqstate}
P\left( \rho \right) =K\rho ^{2},
\end{equation}
with
\begin{equation}
K=\frac{2\pi \hbar ^{2}a}{m^{3}} =0.1856\times 10^5 \left(\frac{a}{1\;{\rm fm}}\right)\left(\frac{m}{2m_n}\right)^{-3},
\end{equation}
where $m_n=1.6749\times 10^{-24}$ g is the mass of the neutron\footnote{In adopting this scaling, we have in mind the possibility that neutrons in the core of neutron stars form the equivalent of Cooper pairs and behave as bosons of mass $2m_n$. This means that we treat the core of neutron stars as a superfluid (see Section \ref{sect8} for additional comments). However, our study may be valid in other circumstances so that we leave the mass $m$ unspecified.}.

Therefore, the equation of state of the Bose-Einstein condensate is a
polytrope with index $n=1$. In this case the solution of the Lane-Emden
equation can be obtained in an analytical form, and the solution satisfying
the boundary condition $\theta \left( 0\right) =1$ is \cite{Ch57}
\begin{equation}
\theta \left( \xi \right) =\frac{\sin \xi }{\xi }.
\end{equation}

The radius of the star is defined by the condition $\theta \left( \xi
_{1}\right) =0$, giving $\xi _1=\pi $. Therefore the radius $R$ of the
Bose-Einstein condensate is given by
\begin{equation}\label{rxj}
R=\pi \sqrt{\frac{\hbar ^{2}a}{Gm^{3}}}=6.61\left(
\frac{a}{1\;{\rm fm}}\right) ^{1/2}\left(\frac{ m}{2m_{n}}\right) ^{-3/2}{\rm km}.
\end{equation}
For $m=2m_{n}$ and $a=1$ fm the radius of the condensate is
$R\approx 7$ km. The radius of the gravitationally bounded
Bose-Einstein condensate is independent on the central density and
on the mass of the star, and depends only on the physical
characteristics of the condensate.

The mass of the star is obtained as
\begin{equation}
\label{mrho}
M=4\pi ^{2}\left( \frac{\hbar
^{2}a}{Gm^{3}}\right) ^{3/2}\rho _{c},
\end{equation}
where we have used $\left| \theta ^{\prime }\left( \xi _{1}\right)
\right| =1/\pi $. As a function of the central density and of the
coherent scattering length $a$ we obtain for the mass of the
Bose-Einstein condensate star with quartic non-linearity the
expression
\begin{equation}\label{Newtmass}
M=1.84\left( \frac{\rho _{c}}{10^{16}\;{\rm g\;cm^{-3}}}\right) \left(
\frac{a}{1\;{\rm fm}}\right) ^{3/2}\left( \frac{m}{2m_{n}}\right) ^{-9/2}M_{\odot }.
\end{equation}
For $m=2m_{n}$, $a=1\, {\rm fm}$ and $\rho_c=5\times 10^{15}\; {\rm g}/{\rm cm}^3$,  the mass of the condensate is $M\approx 0.92\, M_\odot$. However, this mass may be larger than the maximum mass allowed by general relativity (see below), and for a correct determination of the maximum mass of BEC stars we cannot ignore the effects induced by the space-time curvature.

The mass of the static condensate can be expressed in terms of the radius and central density by
\begin{equation}
M=\frac{4}{\pi }\rho _{c}R^{3},
\end{equation}
which shows that the mean density of the star $\overline{\rho}=3M/4\pi R^{3}$ can be obtained from the central density
of the condensate by the relation $\overline{\rho}=3\rho _{c}/\pi ^{2}$.

With respect to a scaling of the parameters $m$, $a$ and $\rho
_{c}$ of the form $m\rightarrow \alpha _{1}m$, $a\rightarrow
\alpha _{2}a$, $\rho _{c}\rightarrow \alpha _{3}\rho _{c}$, the
radius and the mass of the condensate have the following scaling
properties:
\begin{equation}\label{scal}
R\rightarrow \alpha _{1}^{-3/2}\alpha _{2}^{1/2}R,M\rightarrow
\alpha _{1}^{-9/2}\alpha _{2}^{3/2}\alpha _{3}M.
\end{equation}

\subsection{Slowly rotating Bose-Einstein Condensate stars}

The case of slowly rotating Bose-Einstein condensates can also be
straightforwardly analyzed, by taking into account the fact that the
condensate obeys a polytropic equation of state. The study of the slowly
rotating polytropes was performed in detail in \cite{Ch33}.

The Lane-Emden equation for a rotating Bose-Einstein condensate is
\begin{equation}
\frac{1}{\xi ^{2}}\frac{\partial }{\partial \xi }\left( \xi ^{2}\frac{\partial \Theta }{\partial \xi }\right) +\frac{1}{\xi ^{2}}\frac{\partial }{\partial \mu }\left[ \left( 1-\mu ^{2}\right) \frac{\partial \Theta }{\partial \mu }\right] =-\Theta ^{n}+\Omega ,
\end{equation}
where $\mu =\cos \theta $ and $\Omega =\omega ^{2}/2\pi G\rho _{c}$.
The volume $V_{\omega }$ and the mass $M_{\omega }$ of the condensate in slow
rotation are given in the first order in $\Omega $ by
\begin{equation}
\label{vrot}
V_{\omega }=V_0\left[ 1+\frac{3\psi _{0}\left( \xi
_{1}\right) }{\xi _{1}\left| \theta ^{\prime }\left( \xi _{1}\right) \right|
}\Omega \right],
\end{equation}
and
\begin{eqnarray}
\label{mrot}
M_{\omega }=M_0\left[ 1+\frac{\xi _{1}/3-
\psi' _{0}\left(\xi _{1}\right) }{\left| \theta ^{\prime }\left( \xi _{1}\right) \right| }\Omega \right] ,
\end{eqnarray}
respectively. In these expressions, $M_0$ is given by Eq. (\ref{massn}) and $V_0=(4/3)\pi R_0^3$ where $R_0$ is given by Eq. (\ref{radiusn}). The values of the function $\psi _{0}$ are tabulated in \cite
{Ch33}. Equations (\ref{vrot}) and (\ref{mrot}) represent the mass and volume relations for two stars with equal central density, one rotating with an angular velocity $\omega$ and the other non-rotating.

In the case of Bose-Einstein condensates with quartic
non-linearity, corresponding to a polytropic index $n=1$, the Lane-Emden equation can be integrated exactly
(yielding $\theta(\xi)=\sin\xi/\xi$ and  $\psi_0(\xi)=1-\sin\xi/\xi$),  giving for the volume
$V_{\omega }$ and mass $M_{\omega }$ of the rotating condensate
the following simple relations
\begin{equation}
V_{\omega }=V_0\left( 1+3\Omega \right),
\end{equation}
\begin{equation}
M_{\omega }=M_0\left[ 1+\left( \frac{\pi ^{2}}{3}-1\right) \Omega \right] .
\end{equation}

\section{Maximum mass of the static Newtonian Bose-Einstein condensate
stars: Qualitative treatment}\label{sect5}

The numerical values of the basic parameters (mass and radius) of
the condensed object sensitively depend on the mass $m$ of the
particle, on the scattering length $a$, and on the central
density $\rho _{c}$: $R=R(a,m)$, $M=M\left(a,m,\rho
_{c}\right)$. Of course, in general, the values of the mass and
radius of the gravitational condensate depend on the adopted model
for the non-linearity.

The scattering length $a$ is defined as the zero-energy limit of
the scattering amplitude $f$, and it can be related to the particle scattering cross section $\sigma $ by the relation $\sigma =4\pi a^2$ \cite{Da99}.  On the other hand, the
notion that particles like, for example, the quarks, retain their
usual properties and interactions at the very high densities in
the neutron stars may not be viable \cite{Gl00, Ch92}. In our
calculations we use a ``hard" core approximation of the potential.
Therefore we accept that at high densities the ``hard" core
potential is in the QCD range of $1$ fm and the allowed values of
the scattering length $a$ may generally be in the interval
$0.5\;{\rm fm}\leq a<1-2$ fm, corresponding to a scattering cross section of around 1 mb.

The transition temperature to a Bose Einstein Condensate of dense matter can be written as
\begin{eqnarray}
T_{c}&=&\frac{2\pi \hbar ^2}{\zeta ^{2/3}k_B}\frac{\rho ^{2/3}}{m^{5/3}}=
1.650\times 10^{12}\nonumber\\
&&\times\left(\frac{\rho }{10^{16}\;{\rm g/cm ^3}}\right)^{2/3}\left(\frac{m}{2m_n}\right)^{-5/3}\;{\rm K}.
\end{eqnarray}
Neutron stars are born
with interior temperatures of the order of $2 - 5\times 10^{11}$ K, but they rapidly cool down via neutrino emission to temperatures of less than $10^{10}$ K within minutes. Also strange matter, pion condensates, $\lambda $ hyperons, $\delta $ isobars, or free quark matter might form under the initial thermal conditions prevailing in the very young neutron star. Hence a condensation process can take place in the very early stages of stellar evolution. If the core is composed of only ``ordinary" matter (neutrons, protons, and electrons), then when the temperature drops below about $10^9$ K all particles are degenerate. We expect that after a hundred years or so the core will become superfluid \cite{Gl00}, and this may also favor the possibility of a Bose-Einstein Condensation through the BCS-BEC crossover.

Restriction on the maximum
central density and maximum mass of the condensate quartic non-linearity can be obtained from the study of the speed of sound, defined as $c_{s}^{2}=\partial P/\partial \rho $. With the use of Eq.~(\ref{eqstate}) we
obtain $c_{s}^{2}=2K\rho $. The causality condition implies $c_{s}\leq c$, where $c$ is the speed of light.

By introducing the dimensionless parameter $\kappa $, defined as
\begin{equation}\label{kappa}
\kappa=\left(\frac{a}{1\;{\rm fm}}\right)^{1/2}\left(\frac{m}{2m_n}\right)^{-3/2},
\end{equation}
the causality condition gives the following upper bound for the central density of
the condensate:
\begin{equation}\label{bound3}
\rho _{c}\leq \frac{m^{3}c^{2}}{4\pi a\hbar ^{2}}=2.42\times
10^{16}\, \kappa ^{-2}\;{\rm g/cm}^{3}.
\end{equation}
With the use of Eqs.~(\ref{bound3}) and (\ref{mrho}) we obtain the following
restriction on the maximum mass of the Bose-Einstein condensate
with quartic non-linearity:
\begin{equation}
{M}\leq \pi \frac{\hbar c^{2} \sqrt{a}}{\left(G m \right)
^{3/2}}=4.46\, \kappa\, M_\odot.
\end{equation}
For $m=2m_{n}$ we obtain the condition $\rho _{c}\leq \left[2.42
/a({\rm fm})\right] \times 10^{16}$ g/cm$^{3}$. By taking into
account that for this range of high densities a physically
reasonable value for the scattering length is $a\approx 1$ fm, we obtain the
restriction on the maximum mass of the Bose-Einstein condensate
star from the causality condition as $M\leq 4.46M_{\odot }$.

A stronger bound on the central density can be derived from the
condition that the radius of the star $R$ must be greater than the
Schwarzschild
radius $R_{S}=2GM/c^{2}$, $R\ge R_{S}$. For a Bose-Einstein condensate star $R_{S}$ can be expressed as a function of the central density and
of the radius as $R_{S}=8GR^{3}\rho _{c}/\pi c^{2}$. Then the
condition of stability against gravitational collapse gives
\begin{equation}
\rho _{c}\le \frac{m^{3}c^{2}}{8\pi a\hbar ^{2}}=1.21\times
10^{16}\,\kappa ^{-2}\;{\rm g/cm}^{3},
\end{equation}
a relation which for the condensate star with $m=2m_n$ and $a=1$ fm
leads to the constraint $\rho _{c}\le 1.21 \times 10^{16}$
g/cm$^{3}$. The constraint on the maximum mass for the stellar type
Bose-Einstein condensate can be formulated as
\begin{equation}\label{bound}
M\le \frac{\pi}{2}\frac{\hbar c^{2} \sqrt{a}}{\left( Gm\right)
^{3/2}}=2.23\, \kappa\, M_{\odot}.
\end{equation}
With $a=1$ fm and $m=2m_n$ we obtain for the maximum mass of the
Bose-Einstein condensate star the restriction $M\leq 2.23M_{\odot
}$.


For the $n=1$ polytrope the radius of the star is independent on
the central density. Generally, one may consider $a$ as a free
parameter, which must be constrained by the physics of the nuclear
interactions taking place in the system. However, due to the
possible dependence of the free scattering length $a$ on the mass
density, in the case of Bose-Einstein condensates there may be
(indirect) dependence of the radius on the central density of the
star.

Finally, we would like to point out that the estimates on the maximum mass obtained in the present Section are qualitative with respect to the numerical factors, and more precise values of the maximum mass of the BEC stars will be obtained in the next Section by using a full general relativistic approach.

\section{General relativistic Bose-Einstein condensate stars}\label{sect6}

In the previous Sections we have considered the gravitationally
bounded Bose-Einstein condensate stars in the framework of
Newtonian gravity. General relativistic effects may change the
physical properties of compact objects in both a qualitative and
quantitative way. For example, general relativity imposes a strict
limit on the maximum mass of a stable compact astrophysical
object, a feature which is missing for classical Newtonian stars.
Therefore the study of the general relativistic Bose-Einstein
condensates offers a better understanding of their physical
properties. In the present Section, we  study the properties of static general relativistic  Bose-Einstein condensate stars.

\subsection{Static general relativistic BEC stars}

For a static spherically symmetric star, the interior line element is given by
\begin{equation}
ds^{2}=-e^{\nu (r)}dt^{2}+e^{\mu (r)}dr^{2}+r^{2}\left(
d\theta ^{2}+\sin ^{2}\theta d\phi ^{2}\right).
\end{equation}

The structure equations describing a general relativistic compact
star are the mass continuity equation and the
Tolman-Oppenheimer-Volkoff (TOV) equation of standard general
relativity, and they are given by \citep{Gl00}
\begin{equation}
\frac{dM}{dr}=4\pi \rho r^2 , \label{s1}
\end{equation}
\begin{equation}
\frac{dP(r)}{dr}=-\frac{G \left( \rho+P/c^2\right)
\left( 4\pi Pr^{3}/c^{2}+M\right) }{r^{2}\left[ 1-2GM(r)/c^{2}r\right] }.  \label{s2}
\end{equation}
These equations extend the classical condition of hydrostatic
equilibrium for a self-gravitating gas to the context of general
relativity. We have written the energy density as $\epsilon=\rho
c^2$.
The system of equations (\ref{s1})-(\ref{s2}) must be closed by
choosing the equation of state for the thermodynamic pressure of
the matter inside the star,
\begin{equation}
P=P\left( \rho \right) .
\end{equation}

At the center of the star, the mass must satisfy the boundary
condition
\begin{equation}
M(0)=0.
\end{equation}

For the thermodynamic pressure $P$ we assume that it vanishes on
the surface, $P(R)=0$.

The exterior of the Bose-Einstein condensate star is characterized
by the Schwarzschild metric, describing the vacuum outside the
star, and given by \citep{Gl00}
\begin{equation}
\left( e^{\nu }\right) ^{ext}=\left( e^{-\mu }\right) ^{ext}=1-\frac{2GM}{c^2r},\qquad r\geq R.
\end{equation}

The interior solution must match with the exterior solution on the
vacuum boundary of the star.

\subsection{Maximum mass of non-relativistic BECs with short-range interaction: $n=1$ polytropes}

We assume that, in general relativity, the BEC can still be
described by the non-relativistic equation of state
\begin{equation}\label{mm3}
P=K\rho ^2, \qquad {\rm with}\qquad  K=\frac{2\pi a\hbar^2}{m^3},
\end{equation}
corresponding to a polytropic equation of state,  with polytropic index $n=1$. The theory of polytropic
fluid spheres in general relativity has been developed by Tooper
\cite{tooper}, and we shall use his formalism and notations. Therefore,
we set
\begin{equation}\label{mm4}
\rho =\rho_c \theta, \quad p=K\rho_c^2\theta^2, \quad \sigma=\frac{K\rho_c}{c^2},
\end{equation}
\begin{equation}\label{mm5}
r=\frac{\xi}{A}, \quad M(r)=\frac{4\pi\rho_c}{A^3}v(\xi), \quad A=\left (\frac{2\pi G}{K}\right )^{1/2},
\end{equation}
where $\rho_c$ is the central density. In terms of these variables, the TOV equation and the mass continuity equation become
\begin{equation}\label{mm6}
\frac{d\theta}{d\xi}=-\frac{(1+\sigma\theta)(v+\sigma\xi^3\theta^2)}{\xi^2(1-4\sigma v/\xi)},
\end{equation}
\begin{equation}\label{mm7}
\frac{dv}{d\xi}=\theta \xi^2.
\end{equation}
For a given value of the relativity parameter $\sigma$, they have to be solved with the initial condition $\theta(0)=1$ and $v(0)=0$. Since $v\sim \xi^3$ as $\xi\rightarrow 0$, it is clear that  $\theta'(0)=0$. On the other hand, the density vanishes at the first zero $\xi_1$ of $\theta$:
\begin{equation}\label{mm8}
\theta(\xi_1)=0.
\end{equation}
This determines the boundary of the sphere. In the non-relativistic limit $\sigma\rightarrow 0$, the system (\ref{mm6})-(\ref{mm7}) reduces to the Lane-Emden equation (\ref{laneemden}) with $n=1$.

From the foregoing relations, we find that the radius, the mass and the central density  of the configuration are given by
\begin{equation}\label{mm9}
R=\xi_1R_*, \quad M=2\sigma v(\xi_1)M_*, \quad \rho_{c}=\sigma \rho_*,
\end{equation}
 where the scaling parameters $R_*$, $M_*$ and $\rho _*$ can be expressed in terms of the fundamental constants and the parameter $\kappa $ as
\begin{equation}\label{mm10}
R_*=\left (\frac{a\hbar^2}{Gm^3}\right )^{1/2}=2.106\, \kappa\;{\rm km},
\end{equation}
\begin{equation}\label{mm101}
{M_*}=\frac{\hbar c^2\sqrt{a}}{(Gm)^{3/2}}=1.420\, \kappa \, {M_{\odot}},
\end{equation}
\begin{equation}\label{mm102}
\rho_*=\frac{m^3c^2}{2\pi a\hbar^2}=4.846\times 10^{16}\, \kappa ^{-2}\;{\rm g/cm^3}.
\end{equation}
We note that the expression of the scaled radius $R_*$ is the same as in the Newtonian regime (in particular it is independent on $c$), while the scaling of the mass and of the density are due to relativistic effects. By varying $\sigma$ from $0$ to $+\infty$, we obtain the series of equilibria in the form $M(\rho_c)$, $R(\rho_c)$ and $M(R)$.

The velocity of sound is $c_{s}^2=P'(\rho)=2K\rho$. The condition that the velocity of sound at the center of the configuration (where it achieves its largest value) is smaller than the velocity of light can be expressed as $2K\rho_c\le c^2$, or equivalently as $\sigma\le\sigma_s$ with
\begin{equation}\label{mm11}
\sigma_{s}=\frac{1}{2}.
\end{equation}
The values of $\xi_1$ and $v(\xi_1)$ at this point have been
tabulated by Tooper (and confirmed by our numerical study):
\begin{equation}\label{mm12}
\xi_1=1.801, \quad v(\xi_1)=0.4981.
\end{equation}
The corresponding values of radius, mass and central density are
\begin{equation}\label{mm13}
R_{s}=1.801\left (\frac{a\hbar^2}{Gm^3}\right )^{1/2}=3.790\, \kappa \;{\rm km},
\end{equation}
\begin{equation}\label{mm131}
M_{s}=0.498\frac{\hbar c^2\sqrt{a}}{(Gm)^{3/2}}=0.707\, \kappa \, M_{\odot},
\end{equation}
\begin{equation}\label{mm132}
 (\rho_c)_{s}=\frac{m^3c^2}{4\pi a\hbar^2}=2.423\times 10^{16}\, \kappa ^{-2}\;{\rm g/cm^3}.
\end{equation}
However, it is not granted that the criterion $\sigma>\sigma_s$ is
equivalent to the condition of dynamical instability. The principle of
causality is a necessary, but not a sufficient, condition of stability\footnote{Similarly, in statistical mechanics, the condition that
the specific heat is positive in the canonical ensemble is a
necessary, but not a sufficient, condition of canonical stability \cite{ijmpb}.}. The
condition of dynamical instability  corresponds to the
turning point of mass $dM=0$ and there is no reason why this should be
equivalent to $c_s=c$.  In fact, our numerical study demonstrates that
this is not the case. We find that the maximum mass does not exactly
correspond to the point where the velocity of sound becomes equal to
the velocity of light. In the series of equilibria (parameterized by
the central density $\sigma$), the instability occurs {\it sooner}
than predicted by the criterion (\ref{mm11}). We find indeed that instability
(corresponding to the mass peak) occurs for $\sigma\ge \sigma_{c}$ with
\begin{equation}\label{mm14}
\sigma_{c}=0.42.
\end{equation}
The values of $\xi_1$ and $v(\xi_1)$ at this point are
\begin{equation}\label{mm15}
\xi_1=1.888,\qquad v(\xi_1)=0.5954.
\end{equation}
The corresponding values of radius, mass and central density are
\begin{equation}\label{mm16}
R_{{\rm min}}=1.888\left (\frac{a\hbar^2}{Gm^3}\right )^{1/2}=3.974\, \kappa \;{\rm km},
\end{equation}
\begin{equation}\label{maxmass1}
{M_{{\rm max}}}=0.5001\frac{\hbar c^2\sqrt{a}}{(Gm)^{3/2}}=0.710\, \kappa\, {M_{\odot}},
\end{equation}
\begin{equation}
 (\rho_c)_{{\rm max}}=0.42\frac{m^3c^2}{2\pi a\hbar^2}=2.035\times 10^{16}\, \kappa ^{-2}\;{\rm g/cm^3}.
\end{equation}
We also note that the radius of a BEC star is necessarily smaller than
\begin{equation}\label{rxj2}
R_{\rm max}=\pi \sqrt{\frac{\hbar ^{2}a}{Gm^{3}}}=6.61\, \kappa \, {\rm km},
\end{equation}
corresponding to the Newtonian limit ($\sigma\rightarrow 0$). Therefore, its
value $3.974 \kappa\le R ({\rm km}) \le 6.61\kappa$ is very much constrained.

\begin{figure}[!ht]
\includegraphics[width=0.98\linewidth]{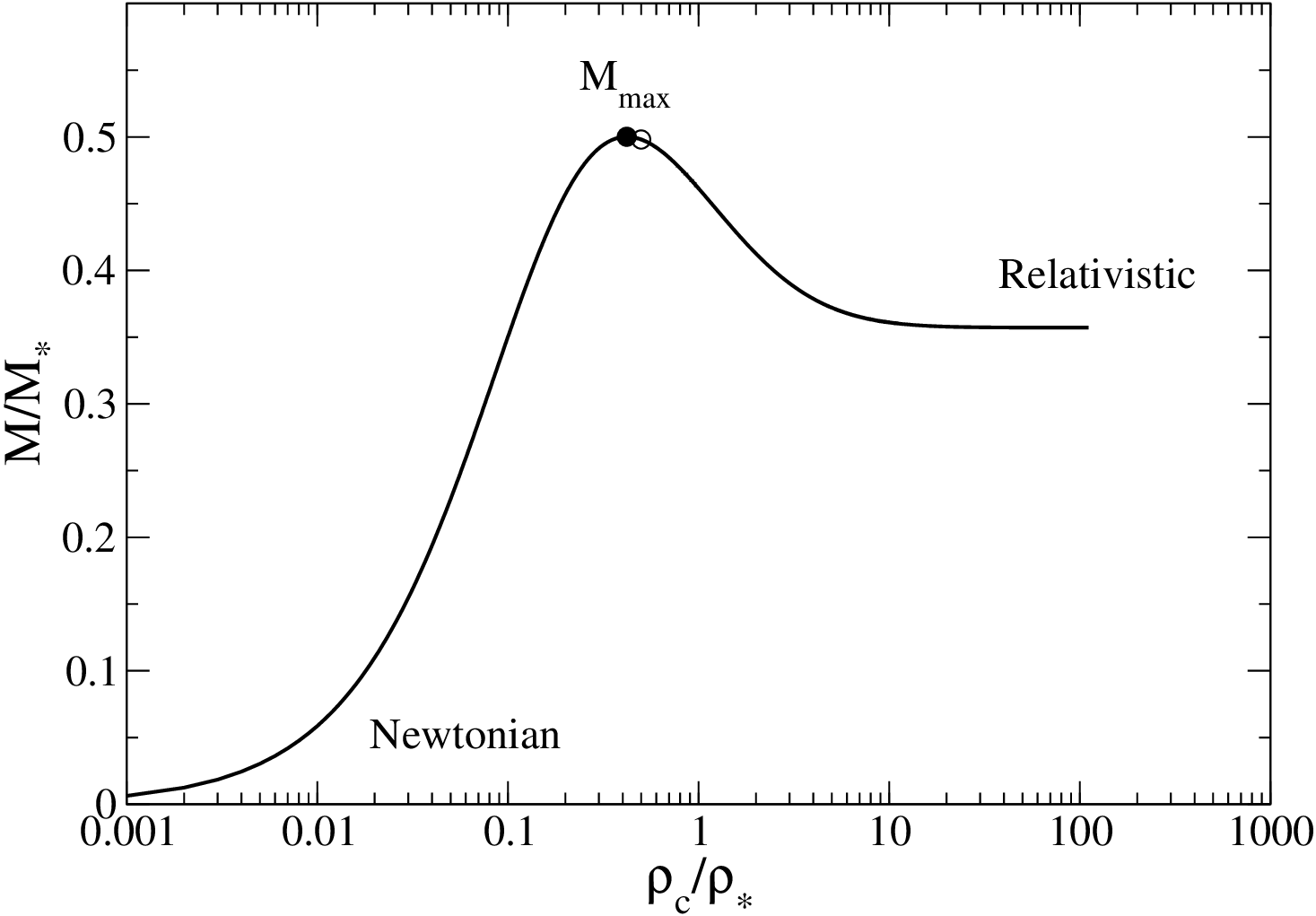}
\caption{Dimensionless mass-central density relation of a relativistic BEC with short-range interactions modeled by a $n=1$  polytrope. There exists a maximum mass $M_{{\rm max}}/M_{*}=0.5001$ (black bullet) at which the system becomes dynamically unstable. This mass does {\it not} coincide with the point (white bullet) at which the velocity of sound becomes equal to the velocity of light (see Fig. \ref{rhoMzoom}). In particular, gravitational instability occurs slightly {\it sooner} than  what is predicted from the criterion based on the velocity of sound (principle of causality). \label{rhoM}}
\end{figure}

\begin{figure}[!ht]
\includegraphics[width=0.98\linewidth]{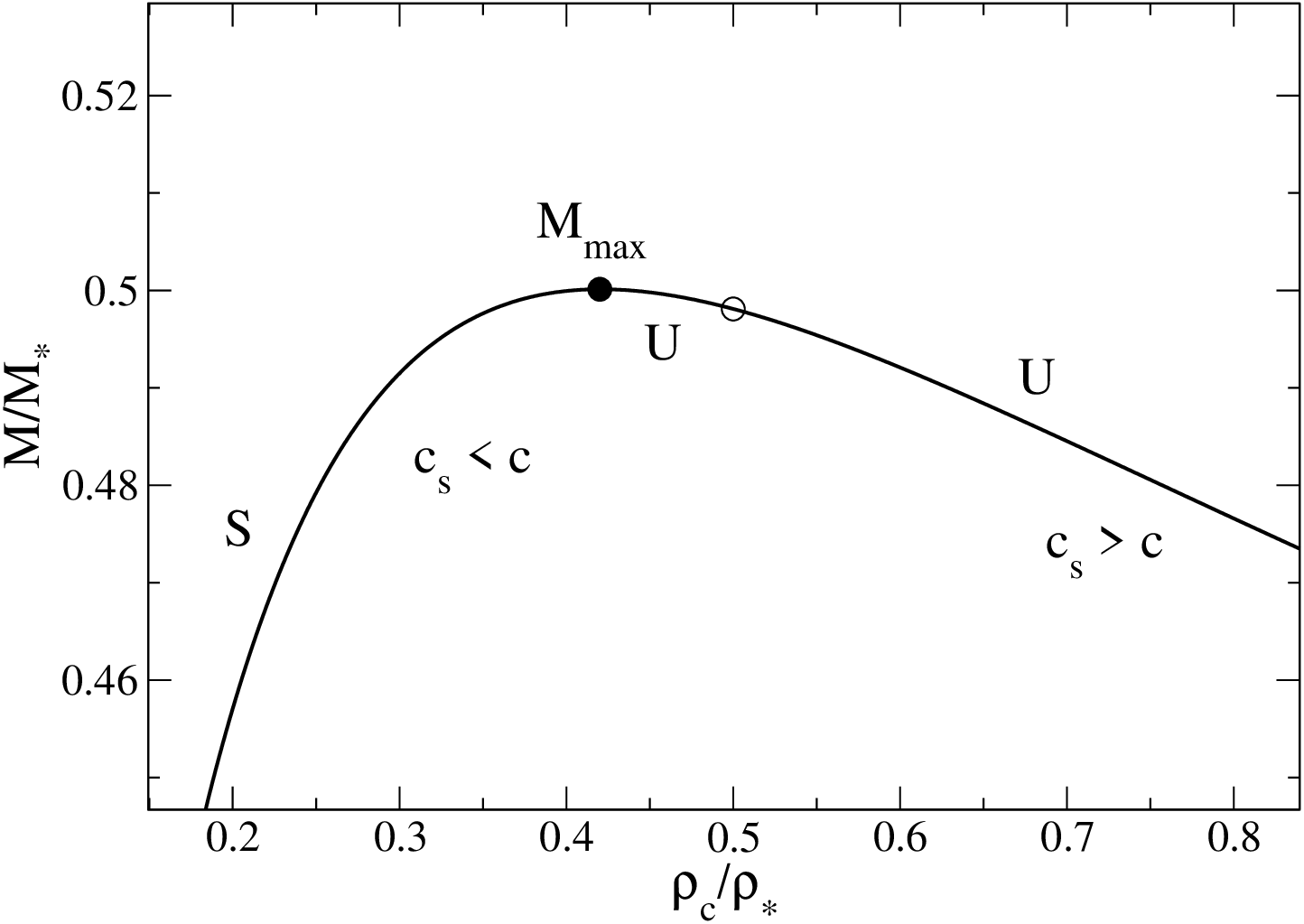}
\caption{Zoom of Fig. \ref{rhoM} showing that the maximum mass does not coincide with the point  at which the velocity of sound becomes equal to the velocity of light. The system is stable (S) before the turning point of mass ($\sigma<\sigma_c$) and unstable (U) after the turning point of mass ($\sigma>\sigma_c$). The velocity of sound is smaller than the velocity of light before the white bullet ($\sigma<\sigma_s$) and larger after that point ($\sigma>\sigma_s$). Therefore, there exists a small region ($\sigma_c<\sigma<\sigma_s$) where the system is unstable although the velocity of sound is smaller than the velocity of light.  \label{rhoMzoom}}
\end{figure}

\begin{figure}[!ht]
\includegraphics[width=0.98\linewidth]{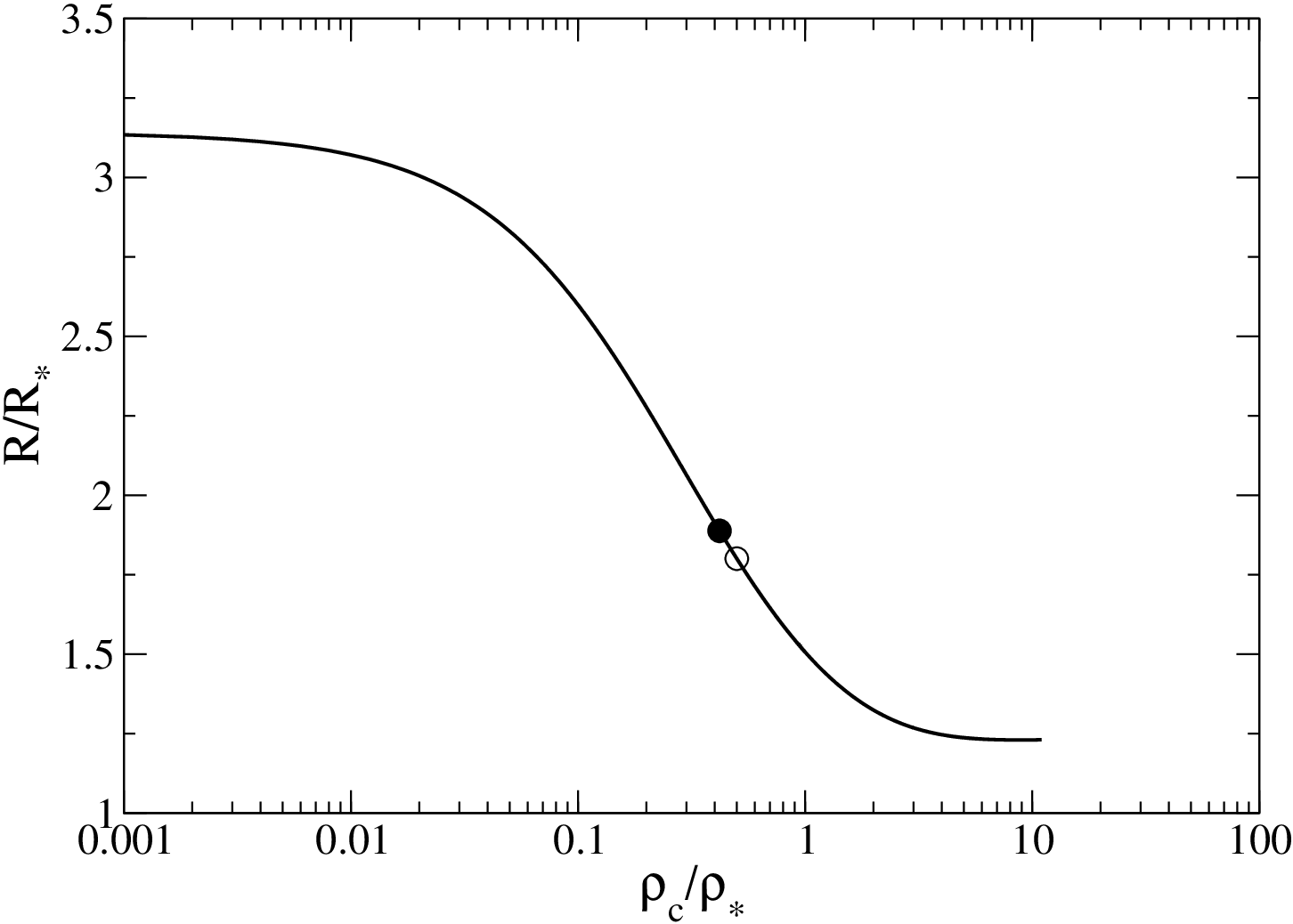}
\caption{Dimensionless radius-central density relation of a relativistic BEC with short-range interactions modeled by a $n=1$  polytrope.   \label{rhoR}}
\end{figure}

\begin{figure}[!ht]
\includegraphics[width=0.98\linewidth]{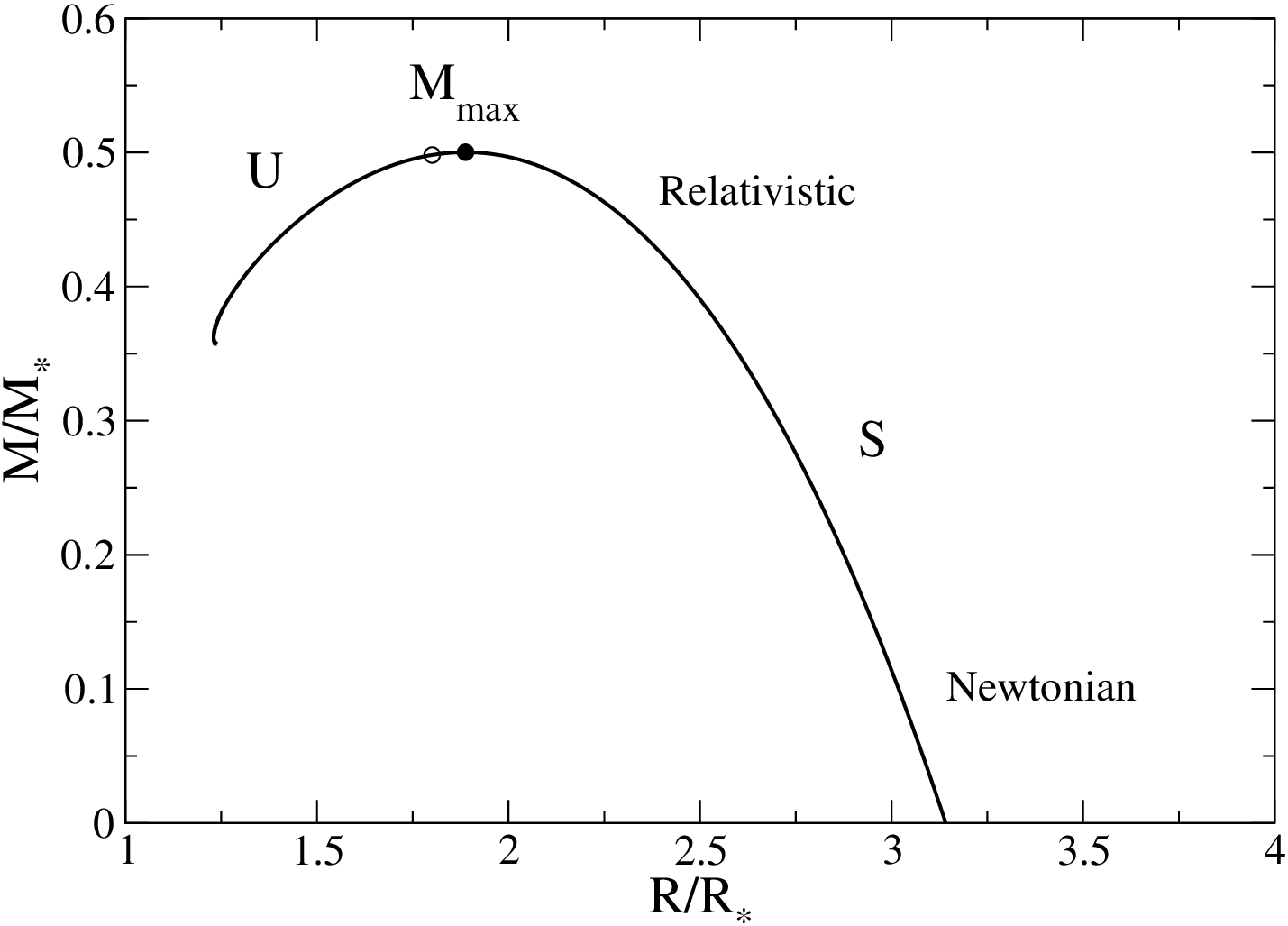}
\caption{Dimensionless mass-radius relation of a relativistic BEC with short-range interactions modeled by a $n=1$  polytrope. There exists a maximum mass $M_{{\rm max}}/M_*=0.5001$ and a minimum radius $R_{{\rm min}}/R_*=1.888$ corresponding to a maximum central density $(\rho_c)_{{\rm max}}=0.42\rho_*$. There also exists a maximum radius $R_{{\rm max}}/R_*=\pi$ corresponding to the Newtonian limit $\sigma\rightarrow 0$.\label{RM}}
\end{figure}

\begin{figure}[!ht]
\includegraphics[width=0.98\linewidth]{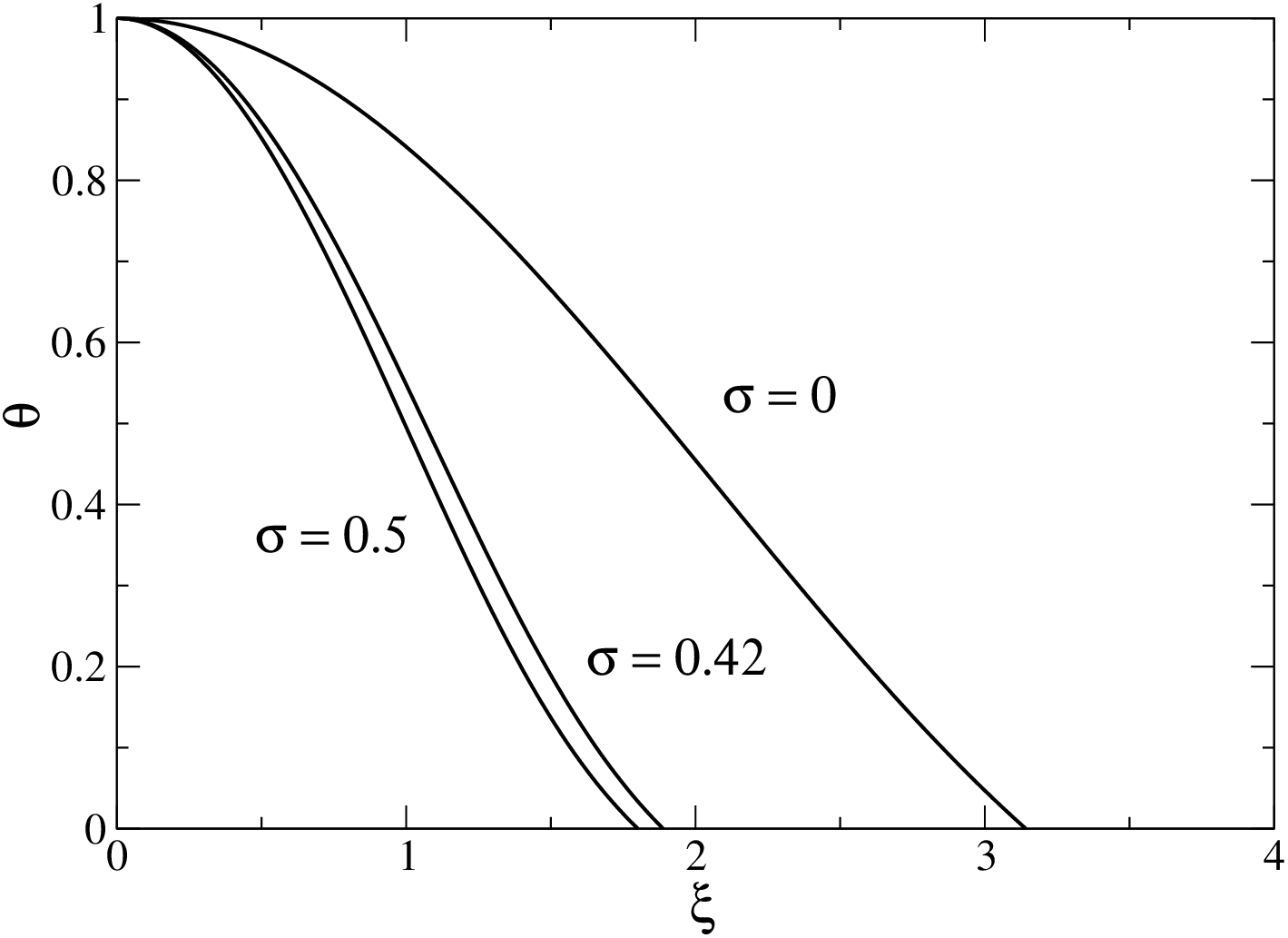}
\caption{Dimensionless density profiles corresponding to $\sigma=0$ (Newtonian), $\sigma=\sigma_c=0.42$ (maximum mass) and $\sigma=\sigma_s=1/2$ (where $c_s=c$). \label{prof}}
\end{figure}

\begin{figure}[!ht]
\includegraphics[width=0.98\linewidth]{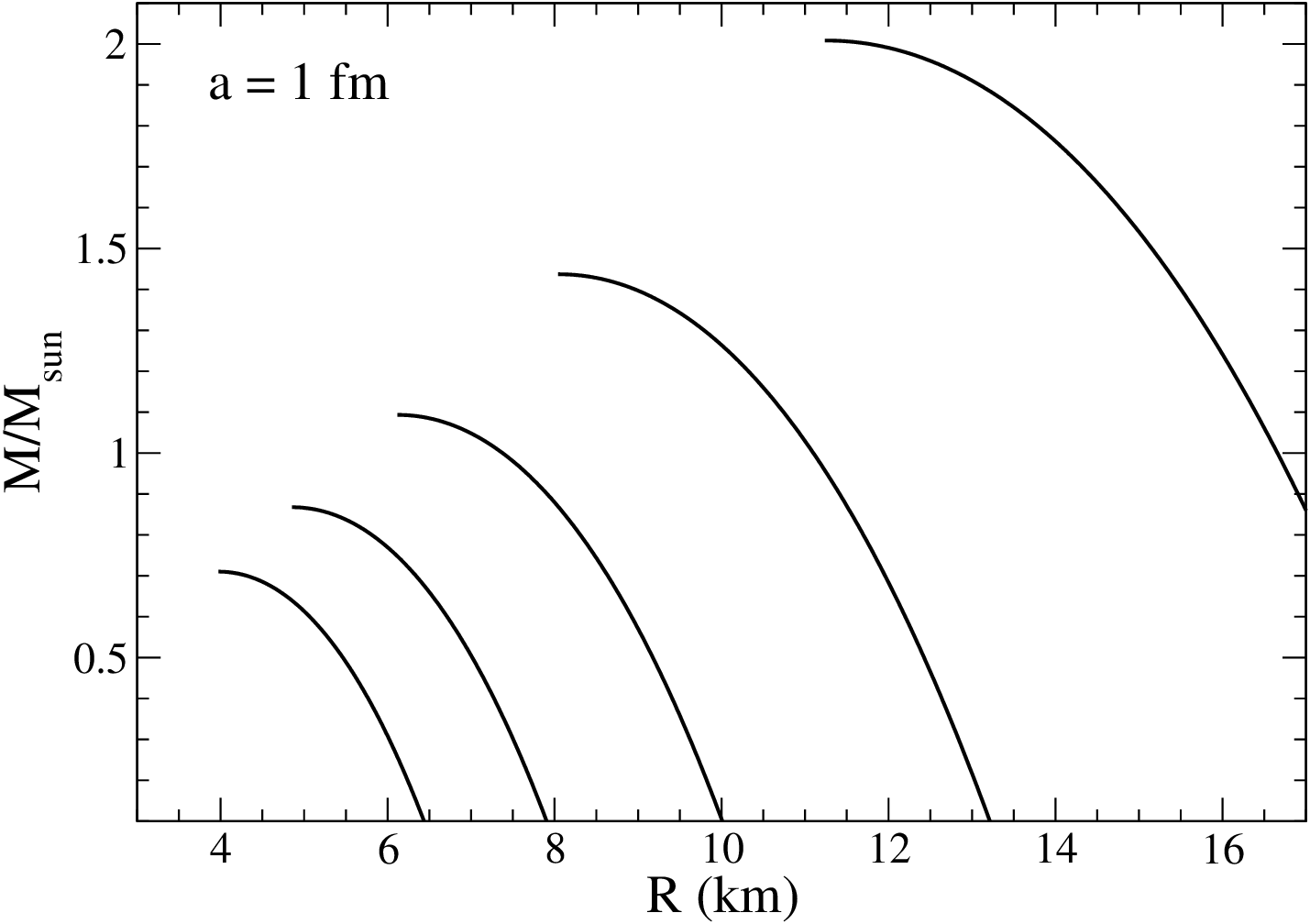}
\caption{Mass-radius dependence for general
relativistic Bose-Einstein condensates with quartic non-linearity for
$a=1$ fm and different values of the mass $m$. From top to bottom:
$m=m_n$, $m=1.25 m_n$, $m=1.5 m_n$, $m=1.75m_n$ and $m=2m_n$. For all
configurations $\rho _{c}\geq
\rho _n$, where $\rho _n=2.026\times 10^{14}\;{\rm g/cm^3}$ is the nuclear density,  and the causality condition $c_s\leq c$ is satisfied.}
\label{FIG8}
\end{figure}

The dimensionless curves giving the mass-central density, radius-central density, mass-radius relations and some density profiles are plotted in Figs. \ref{rhoM}-\ref{prof}. In Fig.~\ref{FIG8}  we present the mass-radius
relation for $a=1$ fm and different values of $m$.

\subsection{Maximum mass of relativistic BECs with short-range interaction}

The previous treatment is approximate because we use the equation of
state (\ref{mm3}) obtained in the non-relativistic regime (i.e. from
the Gross-Pitaevskii equation) but solve the TOV equation
expressing the condition of hydrostatic equilibrium in general
relativity. A fully relativistic approach
based on the Klein-Gordon-Einstein system has been developed by Colpi
{\it et al.}
\cite{colpi}. In order to make the correspondence between BECs with
short-range interactions described by the Gross-Pitaevskii equation
and scalar fields with a $\frac{1}{4}\lambda|\phi|^4$ interaction described by the
Klein-Gordon equation, we set \cite{Chav1}
\begin{equation}
\label{mm17}
\frac{\lambda}{8\pi}\equiv \frac{a}{\lambda_c}=\frac{a m c}{\hbar},
\end{equation}
where $\lambda_c=\hbar/mc$ is the Compton wavelength of the bosons. The equation of state (\ref{mm3}) can be written
\begin{equation}
\label{mm21}
P=K\rho^2,\qquad {\rm with}\qquad K=\frac{\lambda\hbar^3}{4m^4 c}.
\end{equation}
This returns the equation of state obtained by Arbey {\it et al.} \cite{arbey} for a self-interacting scalar field in the non-relativistic regime showing that the relation between $a$ and $\lambda$ given by Eq. (\ref{mm17}) is correct. We note that the parameter  $\lambda$  can be expressed as
\begin{equation}
\label{lambda1}
\frac{\lambda}{8\pi}=9.523\, \frac{a}{1\, {\rm fm}}\frac{m}{2 m_n}.
\end{equation}
We can then use $(\lambda,m)$ instead of $(a,m)$ as independent physical variables. Finally, the dimensionless parameter $\kappa$ can be written
\begin{equation}
\label{lambda2}
\kappa=0.324\,  \left (\frac{\lambda}{8\pi}\right )^{1/2}\left (\frac{2 m_n}{m}\right )^2.
\end{equation}

We can now express the results in terms of $\lambda$. The scaling of the maximum mass is given by
\begin{equation}
\label{mm18}
M_*=\frac{\hbar c^2 \sqrt{a}}{(Gm)^{3/2}}= \sqrt{\frac{\lambda}{8\pi}}\frac{1}{m^2}\left (\frac{\hbar c}{G}\right )^{3/2}=\sqrt{\frac{\lambda}{8\pi}}\frac{M_P^3}{m^2},
\end{equation}
where $M_P=(\hbar c/G)^{1/2}$ is the Planck mass. This is the scaling
of the maximum mass obtained by Colpi {\it et al.} \cite{colpi} for a
self-interacting scalar field. For $\lambda\sim 1$, it is of the order
of the Chandrasekhar mass $\sim M_P^3/m^2$. On the other hand, the
scaling of the minimum radius is given by
\begin{equation}
\label{mm19}
R_*=\left (\frac{a\hbar^2}{Gm^3}\right )^{1/2}=\left (\frac{\lambda\hbar^3}{8\pi Gc}\right )^{1/2}\frac{1}{m^2}=\sqrt{\frac{\lambda}{8\pi}}\frac{M_P}{m}\lambda_c.
\end{equation}
This is the scaling of the minimum radius given by Arbey {\it et al.} \cite{arbey} for a self-interacting scalar field. Finally, the scaling of the maximum density is given by
\begin{equation}
\label{mm20}
\rho_*=\frac{m^3 c^2}{2\pi a\hbar^2}=\frac{4m^4c^3}{\lambda\hbar^3}.
\end{equation}
Now, the maximum mass obtained by Colpi {\it et al.} \cite{colpi} in the fully relativistic regime is
\begin{equation}
\label{mm22}
M_{{\rm max}}=0.22\, \sqrt{\frac{\lambda}{4\pi}}\frac{M_P^3}{m^2}=0.22\sqrt{2}M_*=0.31M_*,
\end{equation}
which is smaller than our previous estimate $M_{\rm max}=0.5001M_*$ based on a non-relativistic equation of state. Therefore, relativistic effects tend to reduce the maximum mass.

Colpi {\it et al.} \cite{colpi} showed that, in the Thomas-Fermi limit, the scalar field becomes equivalent to a fluid with an equation of state
\begin{equation}
\label{mm23}
P=\frac{c^4}{36K}\left\lbrack \left (1+\frac{12K}{c^2}\rho \right )^{1/2}-1\right\rbrack^2,
\end{equation}
where $K$ is given by Eq.~(\ref{mm21}). For $\rho \rightarrow 0$ (low
or moderate densities), we recover the polytropic equation of state
$p=K\rho ^2$ corresponding to a non-relativistic BEC with short-range
interactions. For $\rho \rightarrow +\infty$ (extremely high
densities), we obtain the ultra-relativistic equation of state $p=\rho
c^2/3$, similar to the one describing the core of neutron stars
modeled by the ideal Fermi gas \cite{OV,misner,chavgen}. We know that
a linear equation of state $p=q\rho c^2$ yields damped oscillations of
the mass-central density relation, and a spiral structure of the
mass-radius relation \cite{chavgen}, similarly to the isothermal
equation of state in Newtonian gravity \cite{aaiso}. Therefore, our
BEC model will exhibit this behaviour, just like standard neutron
stars. However, our BEC model differs from standard neutron star
models in that, at low or moderate densities, $p=K\rho^2$ with $K=2\pi
a\hbar^2/m^3$ instead of $p=K'\rho^{5/3}$ with
$K'=(1/5)(3/8\pi)^{2/3}h/m_n^{8/3}$). This implies, in particular, the existence of a maximum radius given by Eq. (\ref{rxj2}), corresponding to the Newtonian limit.

Substituting the equation of state (\ref{mm23}) in the TOV equations, using
\begin{equation}
\label{mm23b}
P'(\rho )=\frac{1}{3}c^2\left\lbrack 1-\frac{1}{\sqrt{1+12K\rho /c^2}}\right\rbrack,
\end{equation}
and introducing the same notations as before,  we obtain
\begin{widetext}
\begin{equation}\label{mm24}
\frac{d\theta}{d\xi}=-\frac{6\left\lbrack \frac{1}{36}(\sqrt{1+12\sigma\theta}-1)^2+\sigma\theta\right\rbrack\left\lbrack v+\frac{\xi^3}{36\sigma}(\sqrt{1+12\sigma\theta}-1)^2\right\rbrack}{\xi^2(1-4\sigma v/\xi)(1-1/\sqrt{1+12\sigma\theta})},
\end{equation}
\begin{equation}\label{mm25}
\frac{dv}{d\xi}=\theta \xi^2,
\end{equation}
\end{widetext}
instead of Eqs. (\ref{mm6})-(\ref{mm7}). If we expand the square roots for $\sigma\ll 1$, we recover Eqs.~(\ref{mm6})-(\ref{mm7}).  However, this is not a uniform expansion and the two equations (\ref{mm24})-(\ref{mm25}) and (\ref{mm6})-(\ref{mm7}) are in fact different even for small values of $\sigma$ (of course, they both reduce to the Lane-Emden equation (\ref{laneemden}) for $\sigma=0$).

\begin{figure}[!ht]
\includegraphics[width=0.98\linewidth]{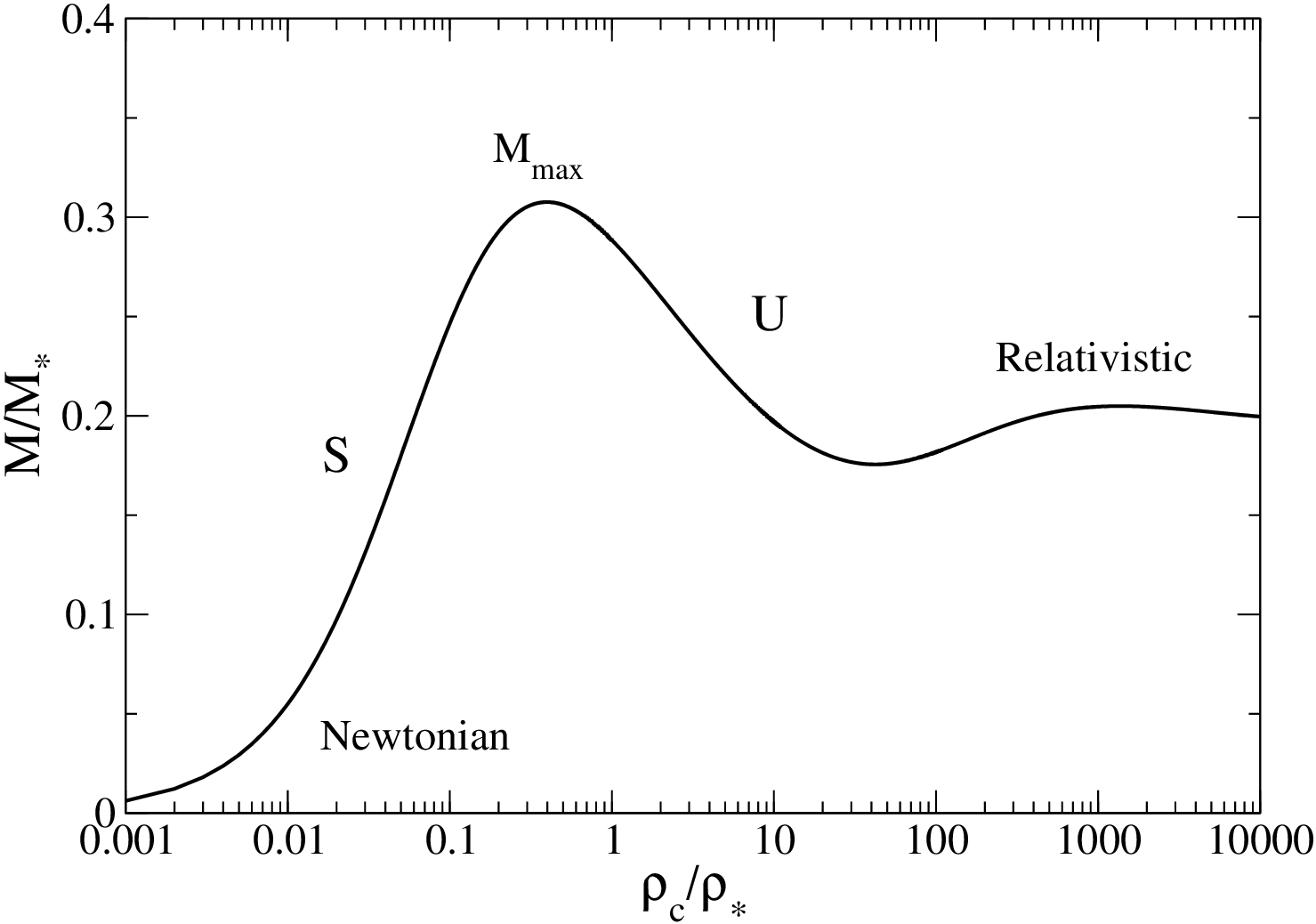}
\caption{Dimensionless mass-central density relation of a relativistic BEC with short-range interactions modeled by the equation of state (\ref{mm23}). There exists a maximum mass $M'_{{\rm max}}/M_{*}=0.3076$ at which the series of equilibria becomes dynamically unstable. The velocity of sound is always smaller than the velocity of light. We note that the  mass-central density relation presents damped oscillations at high densities similarly to standard neutron stars \cite{misner,chavgen}. \label{colpirhoM}}
\end{figure}

\begin{figure}[!ht]
\includegraphics[width=0.98\linewidth]{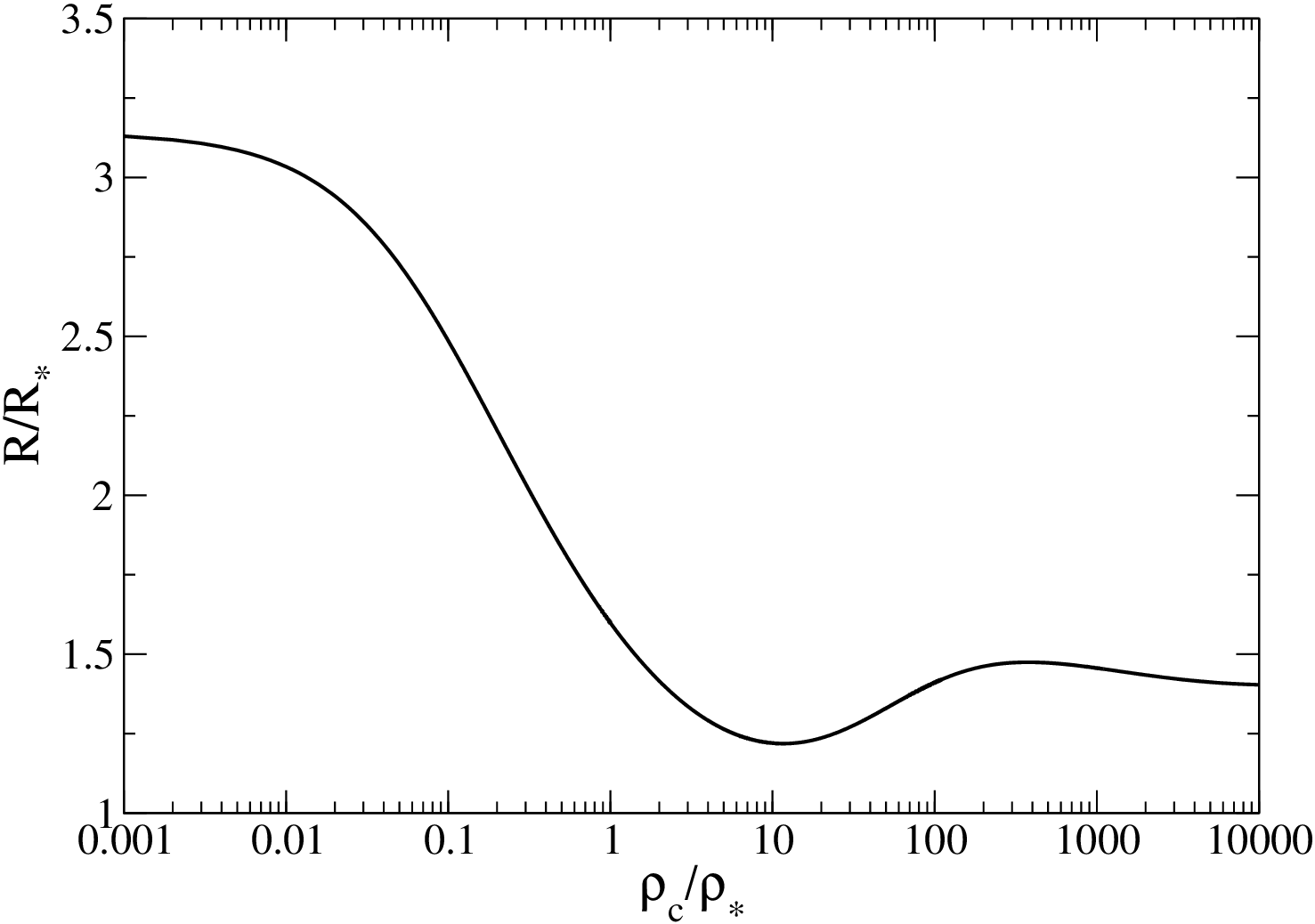}
\caption{Dimensionless radius-central density relation of a relativistic BEC with short-range interactions modeled by the equation of state (\ref{mm23}).   \label{colpirhoR}}
\end{figure}

\begin{figure}[!ht]
\includegraphics[width=0.98\linewidth]{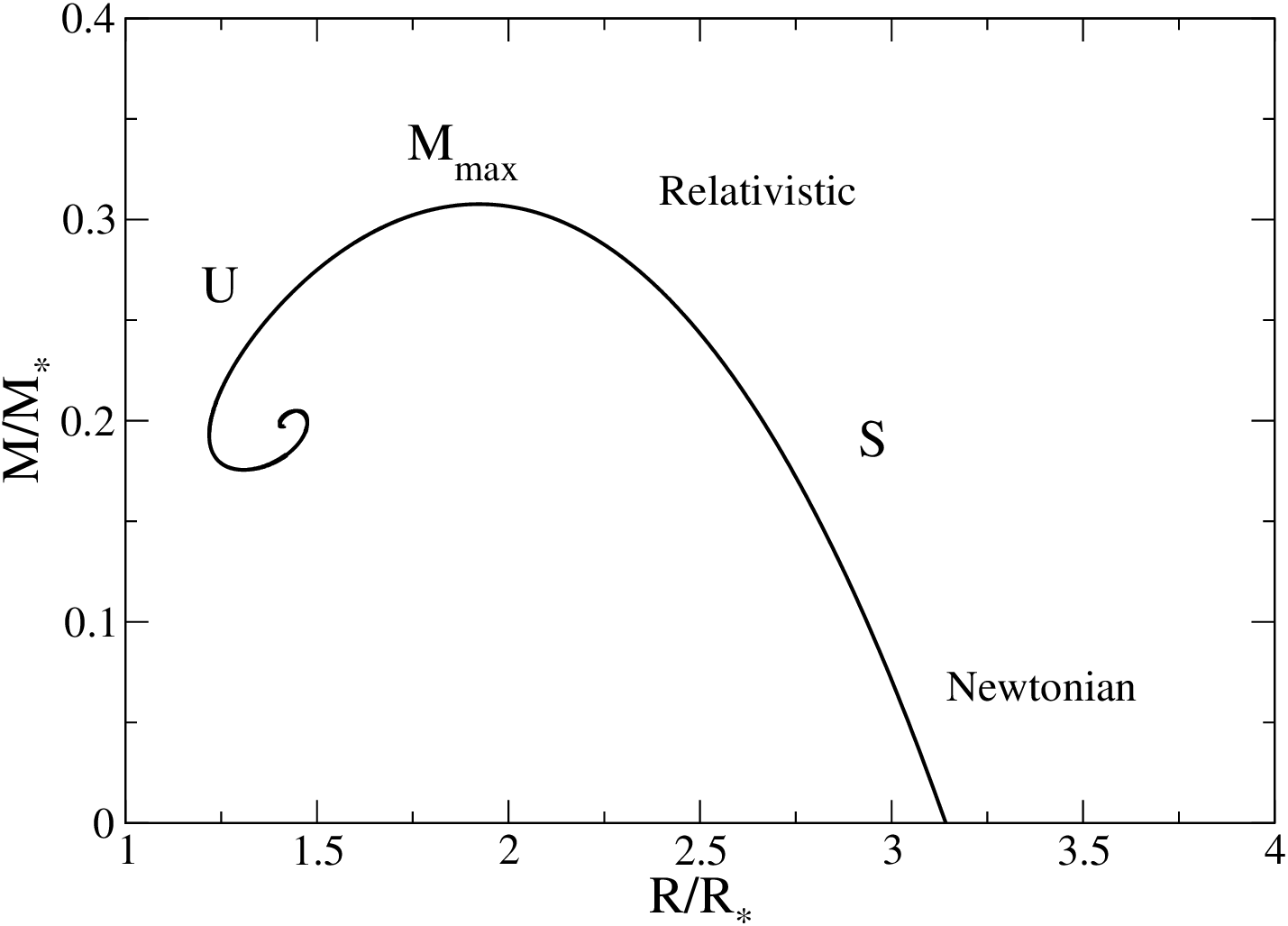}
\caption{Dimensionless mass-radius relation of a relativistic BEC with short-range interactions modeled by the equation of state (\ref{mm23}). The mass-radius relation presents a snail-like structure (spiral) at high densities similarly to standard neutron stars \cite{mt,chavgen}. There exists a maximum mass $M_{{\rm max}}/M_*=0.307$ and a minimum radius $R_{{\rm min}}/R_*=1.923$ corresponding to a maximum central density $(\rho_c)_{{\rm max}}=0.398\rho_*$. There also exists a maximum radius $R_{{\rm max}}/R_*=\pi$ corresponding to the Newtonian limit $\sigma\rightarrow 0$.   \label{colpiRM}}
\end{figure}

\begin{figure}[!ht]
\includegraphics[width=0.98\linewidth]{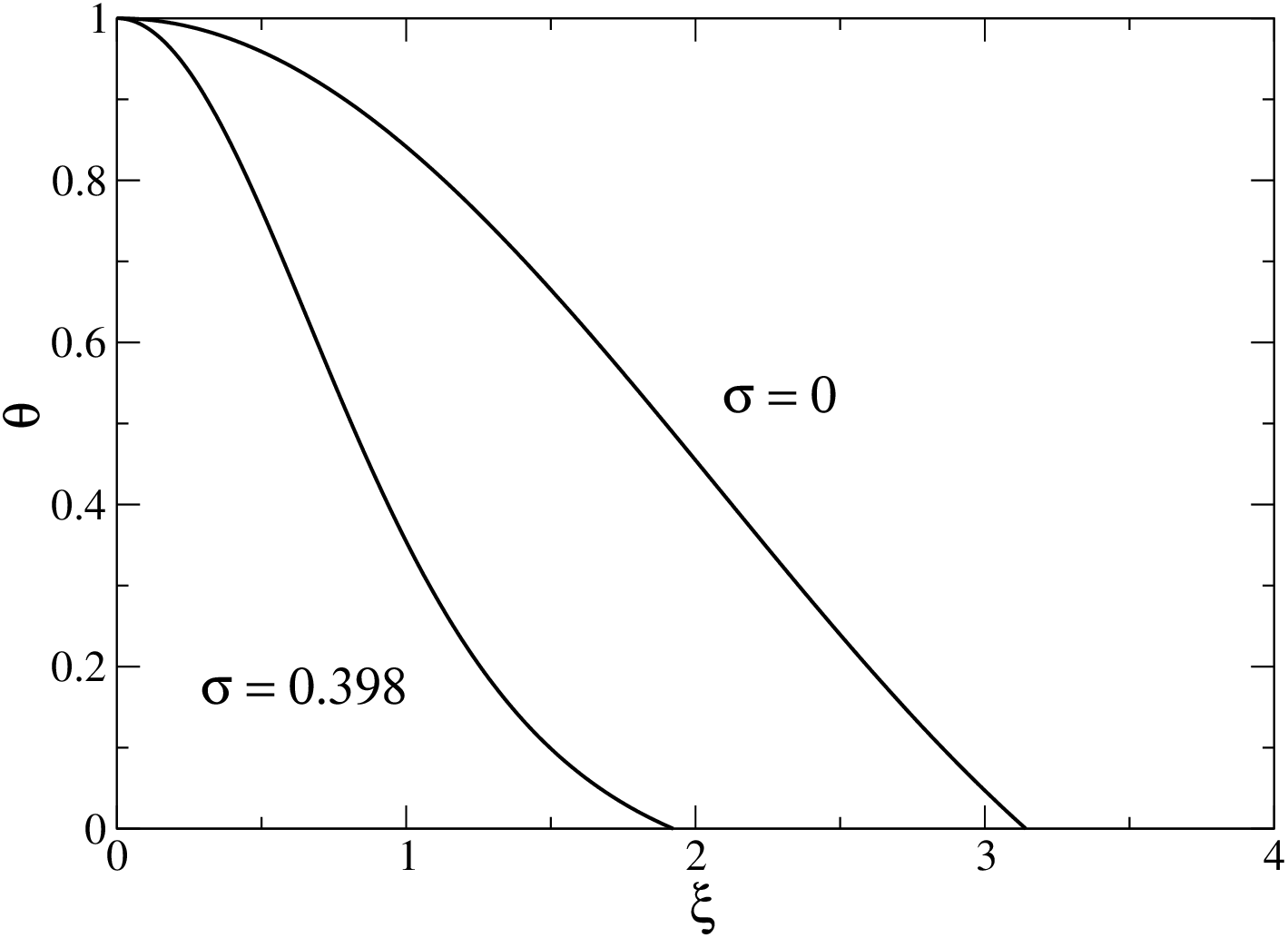}
\caption{Dimensionless density profiles corresponding to $\sigma=0$ (Newtonian) and $\sigma=\sigma'_c=0.398$ (maximum mass).\label{colpiprof}}
\end{figure}

The velocity of sound at the center of the configuration is
\begin{equation}
\label{mm26}
(c_s^2)_0=\frac{1}{3}c^2\left (1-\frac{1}{\sqrt{1+12\sigma}}\right),
\end{equation}
and we always have $(c_s)_0<c$. The series of equilibria becomes unstable after the first mass peak. We find that instability
occurs for $\sigma\ge \sigma'_{c}$ with
\begin{equation}\label{mm27}
\sigma'_{c}=0.398.
\end{equation}
The values of $\xi_1$ and $v(\xi_1)$ at this point are
\begin{equation}\label{mm28}
\xi'_1=1.923, \qquad v(\xi'_1)=0.3865
\end{equation}
The corresponding values of the radius, mass and central density are
\begin{equation}\label{mm29}
R'_{{\rm min}}=1.923\left (\frac{a\hbar^2}{Gm^3}\right )^{1/2}=4.047\, \kappa \;{\rm km},
\end{equation}
\begin{equation}\label{maxmass2}
{M'_{{\rm max}}}=0.307\frac{\hbar c^2\sqrt{a}}{(Gm)^{3/2}}=0.436\, \kappa\, {M_{\odot}},
\end{equation}
and
\begin{equation}
 (\rho_c)'_{{\rm max}}=0.398\frac{m^3c^2}{2\pi a\hbar^2}=1.929\times 10^{16}\, \kappa ^{-2}\;{\rm g/cm^3},
\end{equation}
respectively. The maximum mass $M_{\rm max}=0.307 M_*$  is very close to the one [see Eq. (\ref{mm22})] found
by Colpi {\it et al.} \cite{colpi} by solving the
Klein-Gordon-Einstein equations. This shows the accuracy of the hydrodynamical approach in the TF limit. The dimensionless curves giving the
mass-central density, radius-central density, mass-radius relations
and some density profiles are plotted in Figs.~\ref{colpirhoM}-\ref{colpiprof}.

{\it Remark:} At $T=0$, the first law of thermodynamics takes the form
\begin{equation}
 d\rho=\frac{P/c^2+\rho}{n}dn,
\end{equation}
where $n m$ is the rest-mass density. Integrating this relation with the equation of state (\ref{mm23}), we can obtain the relation $n(\rho)$. For $\rho\rightarrow 0$ (non-relativistic limit), we get $\rho=nm$, leading to $P\sim K\rho^2=K(nm)^2$ corresponding to a polytrope $n=1$. For $\rho\rightarrow +\infty$ (ultra-relativistic limit), we get $\rho\propto (nm)^{4/3}$, leading to $P\sim \rho c^2/3\propto (nm)^{4/3}$ corresponding to a polytrope $n=3$ like for an ultra-relativistic Fermi gas at $T=0$ (standard neutron star). These results are consistent with those obtained by Goodman \cite{goodman}. The proper number of particles is
\begin{equation}
N=\int_0^R n(r)\left\lbrack 1-\frac{2GM(r)}{c^2r}\right \rbrack^{-1/2}\, 4\pi r^2\, dr.
\end{equation}
It can be shown that a general relativistic, spherically symmetric,
gaseous star at $T=0$ is dynamically stable with respect to the
Einstein equations if, and only if, it is a maximum of $N[\rho]$ at
fixed mass $M[\rho]=M$ (see, e.g., \cite{weinberg,chavbh}). The first
order variations $\delta N-\alpha\delta M=0$, where $\alpha$ is a
Lagrange multiplier, yield the TOV equations determining steady state
solutions. The ensemble of these solutions forms the series of
equilibria. Then, using the Poincar\'e theorem \cite{ijmpb}, one can
conclude that the series of equilibria becomes unstable at the first
mass peak and that a new mode of instability appears at each turning
point of mass in the series of equilibria (see \cite{st} for an
alternative derivation of these results). At these points, we have
$\delta N=\delta M=0$ so that the curve $N(M)$ presents {\it cusps} at
each point where $M(\rho_c)$ reaches an extremal value. An
illustration of this behavior is given in Figure 5 of
\cite{chavbh}. These results of dynamical stability for general
relativistic stars are similar to results of dynamical and
thermodynamical stability for Newtonian self-gravitating systems
\cite{aaiso,aaantonov,ijmpb}.

\section{Astrophysical implications}\label{sect7}

One of the most important results in general relativistic astrophysics is the existence of a maximum mass of the neutron stars \cite{OV}. Ultra-dense compact objects may have a stable equilibrium configuration until their mass $M$ is equal to the maximum mass $M_{\rm max}$. By integrating the mass continuity equation and the hydrostatic equilibrium equation for a star made of free, non-interacting, neutron gas, Oppenheimer and Volkoff \cite{OV} have shown that the maximum equilibrium mass is $M_{\rm OV}=0.7M_{\odot}$, with a corresponding radius of the order of $R_{{\rm OV}}=9.6$ km, and a central density of the order of $\rho_c=5\times 10^{15}\,  {\rm g}/{\rm cm}^3$. Using a variational method in which the equation of state was constrained
 to have subluminal sound velocity and to be stable against microscopic collapse, Rhoades and Ruffini \cite{RR},
proved that, in the regions where it is uncertain, the equation
of state that produces the maximum neutron star mass is the
one for which the sound speed is equal to the speed of light, i.e. $p=\rho c^2$. As
a result, they found a maximum neutron star mass $M_{{\rm max}}\approx 3.2 M_{\odot}$,
assuming uncertainty in the equation of state above a fiducial
density $\rho_0 \approx 4.6 \times 10^{14}\;{\rm  g/cm^3}$. More realistic models that take into account the composition of the star and the interaction between neutrons led to values of the maximum mass of the neutron stars in the range $1.5-3.2M_{\odot}$ \cite{Kal}. The main reason for the lack of a better theoretical value of the maximum mass of the neutron star is the poor knowledge of the equation of state of hadronic matter at high densities.

With the use of Eqs.~(\ref{mm10})-(\ref{mm102}) it follows that the scaled mass, radius and central density satisfy the relations
\begin{equation}\label{scal12}
\frac{R_*}{M_*}=\frac{G}{c^2},\quad \rho_* M_*^2=\frac{c^6}{2\pi G^3}, \quad \rho _*R_*^2=\frac{c^2}{2\pi G}.
\end{equation}
With the use of Eqs.~(\ref{mm9}), (\ref{mm14}), (\ref{mm15}) and (\ref{scal12}), we obtain the following radius-mass, central density-mass, and central density-radius relations for the maximally stable non-relativistic BEC configuration,
\begin{equation}
R_{{\rm min}}=\frac{\xi_1}{2\sigma_c v(\xi_1)} \frac{GM_{\rm max}}{c^2}=5.599 \frac{M_{{\rm max}}}{M_{\odot}}\; {\rm  km},
\end{equation}
\begin{eqnarray}
\left(\rho_c\right)_{{\rm max}} &=&4\sigma _c^3v^2\left(\xi _1\right)\frac{c^6}{2\pi G^3M_{\rm max}^2}\nonumber\\
&=&1.026\times 10^{16}\left(\frac{M_{\odot}}{M_{{\rm max}}}\right)^2\;\frac{{\rm g}}{{\rm cm^3}},
\end{eqnarray}
\begin{eqnarray}
\left(\rho _c\right)_{{\rm max}}&=&\sigma _c\xi _1^2\frac{c^2}{2\pi G R_{\rm min}^2}\nonumber\\
&=&3.215\times 10^{15}\left(\frac{10\;{\rm km}}{R_{{\rm min}}}\right)^{2}\;\frac{{\rm g}}{{\rm cm^3}}.
\end{eqnarray}

For the relativistic BEC star, using Eqs. (\ref{mm27}) and (\ref{mm28}), we find
\begin{equation}
R_{{\rm min}}=9.271 \frac{M_{{\rm max}}}{M_{\odot}}  \;{\rm km},
\end{equation}
\begin{equation}
\left(\rho_c \right)_{{\rm max}}=3.682\times 10^{15}\left(\frac{M_{\odot}}{M_{{\rm max}}}\right)^2\;\frac{{\rm g}}{{\rm cm^3}},
\end{equation}
\begin{equation}\label{scalfin}
\left(\rho _c\right)_{{\rm max}}=3.160\times 10^{15}\left(\frac{10\;{\rm km}}{R_{{\rm min}}}\right)^2\;\frac{{\rm g}}{{\rm cm^3}}.
\end{equation}

In the non-relativistic case, the mass-radius ratio of the star can be expressed as
\begin{equation}\label{ratio}
\frac{2GM_{{\rm max}}}{c^2R_{{\rm min}}}=\frac{4\sigma _c v\left(\xi _1\right)}{\xi _1}=0.529,
\end{equation}
while, for the relativistic case, we obtain
\begin{equation}\label{ratio1}
\frac{2GM_{{\rm max}}}{c^2R_{{\rm min}}}=0.319.
\end{equation}
A classical result  by Buchdahl \cite{Bu} shows that for static solutions of the spherically symmetric Einstein-matter systems, the total mass $M$ and the area radius $R$ of the boundary of the body obey the relation $2GM/c^2R\leq 8/9=0.888$, the equality sign corresponding to constant density stars. For BEC stars, Eqs.~(\ref{ratio}) and (\ref{ratio1}) obviously satisfy the Buchdahl inequality for the mass-radius ratio.

We emphasize that the radius $R_{\rm min}$ and the central density $(\rho_c)_{\rm max}$ given by Eqs.~(\ref{scal12})-(\ref{scalfin}) only depend on the mass $M_{\rm max}$. In particular, they do not explicitly depend on the two physical parameters of the model, the scattering length $a$, and the particle mass $m$.
On the other hand, the maximum mass $M_{{\rm max}}$ depends on these two parameters only through their ratio $a/m^3$, or equivalently, through the parameter $\kappa $, and it can be obtained from the relations
\begin{equation}\label{mmaxkappa}
M_{{\rm max}}=0.71\, \kappa\, M_{\odot},\qquad M^{\prime}_{{\rm max}}=0.4368\,\kappa \, M_{\odot},
\end{equation}
respectively. Hence, all physical parameters of the model are determined by the mass $M_{\rm max}$ of the star,  which can be obtained from  observations. Therefore, in the present model we have only {\it one} free parameter $\kappa=\left(a/{\rm fm}\right)^{1/2} (2m_n/m)^{3/2}$. With respect to a scaling of the scattering length and of the particle mass of the form
\begin{equation}
a/{\rm fm}\rightarrow\beta _1\left(a/{\rm fm}\right), m/2m_n\rightarrow\beta _2\left(m/2m_n\right),
\end{equation}
where $\beta_1,\beta_2$ are constants, the parameter $\kappa $ scales as
\begin{equation}
\kappa \rightarrow \beta _1^{1/2}\beta _2^{-3/2}\kappa .
\end{equation}
Since $\kappa=0.324\left(\lambda/8\pi\right)^{1/2} (2m_n/m)^{2}$, we equivalently conclude that the maximum mass $M_{{\rm max}}$ depends on the two parameters ($\lambda,m)$ only through their ratio $\lambda/m^4$.

By assuming that the mass of the star is  $M_{{\rm max}}=2M_{\odot}$, in the non-relativistic case we obtain for the parameters of the star
$(\rho_c)_{{\rm max}}=0.256\times 10^{16}\; {\rm g/cm^3}$ and $R_{{\rm min}}=11.2$ km, respectively, independently
on the values of $a$ and $m$.  On the other hand in this case $\kappa =2.816$. If we take $m=2m_n$, this corresponds to a scattering length $a=7.93$ fm, and a coupling constant $\lambda=1.90\times 10^3$. For the relativistic equation of state, we find
$(\rho_c)_{{\rm max}}=0.091 \times 10^{16}\;{\rm g/cm^3}$ and $R_{\rm min}=18.54$ km. For the parameter $\kappa $ we obtain $\kappa =4.578$. If we take $m=2m_n$, this corresponds to a scattering length $a=21.0$ fm and a coupling constant $\lambda=5.02\times 10^3$.  Therefore if we assume values of $\kappa $ of the order of $\kappa \sim 3$ in the non-relativistic regime, and $\kappa \sim 5$ in the relativistic regime, we obtain stellar objects with physical parameters in the range  $M\sim 2M_{\odot}$, $R\sim 10-20$ km, and $\rho_c\sim 0.3-0.1\times 10^{16}\;{\rm  g/cm^3}$, respectively. The only free parameter in the model, $\kappa $,  uniquely determines the mass $M_{\rm max}$ of the star (or conversely).

It may be of interest to make a connection with the results of Oppenheimer \& Volkoff \cite{OV}. In their model, the mass, the radius, and the central density of the critical configuration are
\begin{equation}
\label{ov1}
M_{\rm OV}=0.376 \left (\frac{\hbar c}{G}\right )^{3/2}\frac{1}{m_n^2}=0.7\, M_{\odot},
\end{equation}
\begin{equation}
\label{ov2}
R_{\rm OV}=9.36  \frac{G M_{OV}}{c^2}=9.6\, {\rm km},
\end{equation}
\begin{equation}
\label{ov3}
(\rho_c)_{\rm OV}=3.92\times 10^{-3} \frac{c^6}{G^3 M_{OV}^2}=5\times 10^{15}\, {\rm g}/{\rm cm}^3.
\end{equation}
In our model, introducing the parameter $\lambda$, the maximum mass is given by
\begin{equation}
\label{ov4}
M_{\rm max}=2\sigma_c v(\xi_1) \sqrt{\frac{\lambda}{8\pi}}\left (\frac{\hbar c}{G}\right )^{3/2}\frac{1}{m^2}.
\end{equation}
If we write the boson mass as $m=k m_n$, we obtain the maximum mass in the form
\begin{equation}
\label{ov5}
M_{{\rm max}}= C \frac{\sqrt{\lambda }}{k^2} M_{{\rm OV}},
\end{equation}
where $C=0.265$ for a non-relativistic equation of state and $C=0.163$
for a relativistic equation of state. Using the previous relations, we
can then easily relate $R_{\rm min}$ and $(\rho_c)_{\rm max}$ to
$R_{\rm OV}$ and $(\rho_c)_{\rm OV}$. Equation (\ref{ov5}) clearly
shows that, with respect to the standard Oppenheimer-Volkoff model, we
have an additional parameter $\lambda $ (the strength of the
self-interaction), which gives the possibility of obtaining higher
values for the maximum mass. Using Eq. (\ref{lambda2}) which becomes
$\kappa=0.258\sqrt{\lambda}/k^2$, we can rewrite Eq. (\ref{ov5}) as
\begin{equation}
\label{ov6}
M_{{\rm max}}= 3.87 \, C \kappa M_{{\rm OV}}.
\end{equation}

Presently, there is conclusive observational evidence from pulsar studies for the existence of neutron stars with
masses significantly greater than $1.5M_{\odot}$ \cite{Lat}. By using the Shapiro time delay to measure the inclination,  the mass of PSR J1614-223048 was recently determined to be  $1.97\pm 0.04 M_{\odot}$ \cite{Dem}. Moreover, a number of X-ray binaries seem to contain high-mass neutron
stars: about $1.9M_{\odot}$ in the case of Vela X-1 and $2.4M_{\odot}$ in the case of 4U 1700-377 \cite{Lat}. Even more intriguing is the case of the  black widow pulsar B1957+20, with a best mass estimate of
about $2.4M_{\odot}$ \cite{black}. This system has both pulsar timing and optical light curve information. B1957+20 is located in an eclipsing binary system, consisting of the 1.6 ms pulsar in a nearly circular 9.17 h period orbit, and  an extremely low mass companion, $M_c\approx  0.03M_{\odot}$. It is believed
that irradiation of the companion by the pulsar strongly heats its cosmic environment to the
point of ablation, leading to a comet-like tail, and a large cloud of plasma. The plasma cloud
is responsible for the eclipsing. The pulsar is literally consuming its companion,
hence the name "black widow". The mass of the companion star has been reduced  to a small fraction of its
original mass \cite{Lat}. On the other hand, a measured mass of $2.4 M_{\odot}$ would be incompatible with hybrid star models containing {\it significant} proportions of exotic matter in the form of hyperons, some forms of Bose condensates, or quark matter \cite{Lat}.

However, the mass and radius of the $2-2.4M_{\odot}$ neutron stars perfectly fit the expected
properties of a Bose-Einstein condensate star. For $\kappa \sim 3$, the mass of a typical general relativistic Bose-Einstein condensate star
 is of the order of two solar masses, with a radius of around 11
km. Therefore, we propose that the recently observed $2-2.4M_{\odot}$ mass neutron stars could be  typical
Bose-Einstein condensate stars.

A last comment may be in order. If we apply the same model (self-gravitating BEC with short-range interactions) to dark matter \cite{dark}, and use the Newtonian approximation (which is valid in this context), the radius of a dark matter halo is given by Eq. (\ref{rxj}), which can be rewritten
\begin{eqnarray}
\label{dm}
R=1.746\times 10^{-2}\left(
\frac{a}{1\;{\rm fm}}\right) ^{1/2}\left(\frac{ m}{1{\rm eV}/c^2}\right) ^{-3/2}{\rm kpc}.\nonumber\\
\end{eqnarray}
Again, we note that the radius $R$ determines the ratio $a/m^3$ or $\lambda/m^4$. Estimating the radius of dark matter halos by  $R=10$ kpc, we obtain $m^3/a=3.049\times 10^{-6}\;({\rm eV}/c^2)^3/{\rm fm}$ and $m^4/\lambda=23.94\; ({\rm eV}/c^2)^4$ \cite{Chav2}.

\section{Discussions and final remarks}\label{sect8}

In the present paper, we have proposed that the core of neutron stars
is a superfluid in which the neutrons form equivalent of Cooper pairs,
so that they act as bosons of mass $2m_n$. Therefore, once the
Bose-Einstein condensation takes place, the neutron star should be
modeled as a self-gravitating BEC star (boson star). In our approach,
we also assume that the bosons have a self-interaction, described by a
scattering length $a$.  The basic properties of the gravitationally
bounded Bose-Einstein condensates have been obtained in both Newtonian
and general relativistic regimes. To obtain the physical
characteristics of the system, we have used the Madelung
representation in which condensates can be modeled by using the
hydrodynamic Euler equations describing a gas whose density and
pressure are related by a barotropic equation of state. For the study
of the Bose-Einstein condensate we have adopted the Thomas-Fermi
approximation, which is valid if the total number of particles $N$
obeys the condition $N\gg R/\pi a$ \cite{Wa01,Chav1,Chav2}, a
condition which can be reformulated, with the use of the mass density
of the condensate, as $\overline{\rho} \gg 3m/4\pi ^{2}R^{2}a$.  This
restriction is obviously satisfied by condensates with densities of
the same order as the nuclear density.

In the physically most interesting case, corresponding to a
quartic non-linearity term in the energy functional, the equation
of state of the Bose-Einstein condensate is that of a polytrope
with polytropic index $n=1$. In this case, the radius
and the mass of the Newtonian stellar condensate can be obtained
in an exact form. In a Bose-Einstein  condensed neutron star the mass $m$ of the particle does not
need to coincide with the neutron mass. For the mass of the condensed particle we have used
an effective value of the order of
$m^*=2m_n$. This value is justified by the high densities in
the neutron star cores, where the process of Bose-Einstein
condensation is most likely to occur via the formation of Cooper pairs. However, we have also explicitly presented the numerical values of the basic physical parameters of the stars for other values of the mass.

General relativistic effects impose strong constraints on the maximum
mass. In the framework of the general relativistic approach one must
numerically integrate the structure equations of the star. In this
way, we obtain a large class of stable astrophysical objects, whose
basic parameters (mass and radius) depend on the particle mass $m$ and
scattering length $a$.  Since the values of $a$ and $m$ are not
well-known, this offers the possibility to obtain a maximum mass for
neutron stars that is larger than the Oppenheimer-Volkoff limit of
$0.7\, M_\odot$, and may be compatible with recent observational
determinations of the masses of some neutron stars. This is possible
because we have two new parameters in our model, the boson mass $m$
and the scattering length $a$, which give additional freedom (although
these parameters should be ultimately determined by fundamental
physics). We have found that the maximum mass $M_{\rm max}(\kappa )$ of
the condensate star, given by Eqs.~(\ref{maxmass1}) or
(\ref{maxmass2}), depends in fact on a {\it single} parameter
$\kappa(a, m)$ which is proportional to the ratio $a/m^3$ (or,
equivalently, to the ratio $\lambda/m^4$). Since the radius $R_{\rm min}$
and the density $(\rho_c)_{\rm max}$ only depend on $M_{\rm max}$, all the
physical properties of the BEC stars are determined by the parameter
$\kappa $.  Condensates with particle masses of the order of two
neutron masses and scattering length of the order of $10-20$ fm
(corresponding to $\kappa\sim 3-5$) have maximum masses of the order
of $2M_{\odot }$, minimum radii in the range of $10-20$ km and maximum
central density $\rho_c\sim 0.3-0.1\, 10^{16}\, {\rm g}/{\rm cm}^3$ in
the non-relativistic and relativistic regimes, respectively.  On the
other hand, for $a=1$ fm, the maximum mass of the condensate varies
between $0.4-0.7 \, M_\odot$ for $m=2m_n$ (corresponding to
$\kappa=1$), between $1-2 \, M_\odot$ for $m=m_n$ (corresponding to
$\kappa=2.8$), and between $10-16 \, M_\odot$ for $m=m_n/4$
(corresponding to $\kappa=22$), a value which may correspond, for
example, to the effective (density dependent) kaon mass $m_K^*$ in the
interior of neutron stars. Kaon condensation may provide an important
example of Bose-Einstein type stellar condensate
\cite{Gl00}. Attraction from nuclear matter could bring down the mass
of the kaon to an effective value of $m_K^*\approx 200 \, {\rm
MeV}/c^2\approx m_n/9.38\, {\rm MeV}/c^2$. From Eq.~(\ref{Newtmass})
it follows that Newtonian kaon condensates with kaon effective mass of
the order of $m_K^*=m_n/10$ could have masses as high as
$10^6M_{\odot}$, and radii of the order of $600$ km [see
Eq. (\ref{rxj})]. Thus, Bose-Einstein condensate stars, formed from
small mass particles, may represent viable candidates for the
super-massive ``black holes" that reside at the galactic
centers. However, general relativistic effects strongly restrict the
value of the maximum mass of super-massive BEC stars. From
Eq.~(\ref{kappa}) it follows that for $m=m_K^*=m_n/10$ we have $\kappa
=89.44$. With the use of Eq.~(\ref{maxmass1}) we obtain a maximum mass
of $M=63.50 \, M_{\odot}$ for the kaon condensate star, with a radius
$R=355$ km [see Eq. (\ref{mm16})]. On the other hand, smaller mass condensed
particles can have significantly higher maximum relativistic
masses. Hence the Bose-Einstein condensation process in the early
universe may have provided the seeds from which super-massive black
holes were eventually formed through accretion of interstellar matter.

Presently,
the mass of the neutron stars can be determined very accurately, and
many of them have masses in the range of $2-2.4$ solar masses, which
are very difficult to explain by the standard neutron matter models,
including those with exotic matter like quarks. However, these mass
values could be very easily explained by our model if neutron stars
can be modeled as BEC stars.

Bose-Einstein condensate stars could have a normal matter crust,
since we expect that the condensation cannot take place at
densities smaller than the nuclear density or quark deconfinement
density. The presence of the thin crust increases the mass and the
radius of the condensate star by a factor of $10\%$ or $17\%$,
respectively. Therefore, the presence of a neutron crust does not
modify significantly the basic physical properties of the star.
Distinguishing between Bose-Einstein condensate stars and
``standard" neutron stars or other type of condensate or quark
stars could be an extremely difficult observational task. Similarly to the case
of quark stars \cite{ChHa03,HaCh05}, we suggest that high energy
radiation processes from the surface of the condensate may provide
some distinctive features allowing a clear differentiation of
these different types of stellar objects.

In a very general approach, one may assume that the masses $m$ of the
particles forming the stellar type condensate are anisotropic, and
they should be described by a mass tensor $m_{ij}$. Such anisotropic
masses are known from condensed matter physics where they are
encountered in effective mass calculations for electrons immersed in a
band structure, in the case of excitons (electron-hole couples held
together by the Coulomb attraction) and in BEC for semiconductors. The
doping structure of the semiconductor and its anisotropies would give
place to an effective mass matrix for the paraexcitons (singlet
excitons) at least in the low momentum approximation \cite{Ch05}. A
different value for the effective mass $m$ may considerably increase
(or decrease) the total mass of the condensate.

A rotating Bose-Einstein condensate may exhibit a very complex
internal structure and dynamics, mainly due to the presence of
vortex lattices. The vortex lattices may evolve kinetically, with
each vortex following the streamline of a quadrupolar flow. The
quadrupolar distortions can lead to a disordering of the vortex
lattice, and to an instability due to  inter-particle
collisions, finite temperature effects or to the quadrupolar
distortions induced by the external potential. On the other hand,
 due to the high
neutrino emissivity, which is significantly enhanced due to the
condensation, kaon condensate stars are very dark objects. Hence their observational detection may prove to be
an extremely difficult task. The possible
astrophysical/observational relevance of these processes will be
considered in a future publication.

\section*{Acknowledgements}

T. H. is supported by a GRF grant of the government of the Hong Kong SAR.


\begin{thebibliography}{99}

\bibitem{Da99}  F. Dalfovo, S. Giorgini, L. P. Pitaevskii,
 and S. Stringari,   Rev. Mod. Phys. {\bf 71}, 463 (1999);  E. A. Cornell and C. E. Wieman, Rev. Mod. Phys. {\bf 74}, 875 (2002);  W. Ketterle,  Rev. Mod. Phys. {\bf74}, 1131 (2002); L. Pitaevskii and
S. Stringari,   Bose-Einstein condensation, Clarendon Press, Oxford (2003);  R. A. Duine and H. T. C. Stoof,
 Phys. Repts. {\bf 396}, 115 (2004).

\bibitem{Pet}   C. J. Pethick and H. Smith,
	Bose-Einstein condensation in dilute gases, Cambridge, Cambridge University Press, (2008).


\bibitem{LaLi80}  L. D. Landau  and E. M. Lifshitz,  Statistical Physics, Pergamon Press, Oxford (1980).

\bibitem{sopik} J. Sopik, C. Sire, P. H. Chavanis, Phys. Rev. E {\bf 74}, 011112 (2006).


\bibitem{Ch05} Q. Chen, J. Stajic, S. Tan,  and K. Levin,
  Phys. Rept. {\bf 412}, 1 (2005).

\bibitem{An95}  M. H. Anderson, J. R.  Ensher,
M. R. Matthews, C. E. Wieman,  and E. A. Cornell,   Science {\bf 269}, 198 (1995).

\bibitem{Br95}  C. C. Bradley, C. A. Sackett,
J. J. Tollett,  and R. G. Hulet, \prl {\bf 75}, 1687 (1995).

\bibitem{Da95}  K. B. Davis, M. O. Mewes, M.
R. Andrews, N. J. van Drutten, D. S. Durfee, D. M. Kurn,  and W. Ketterle, W., \prl {\bf 75}, 3969 (1995).

\bibitem{Bal95}  M. Baldo, U. Lombardo,  and
P. Schuck, \prc {\bf 52}, 975 (1995); H. Stein, A. Schnell, T. Alm,
 and G. Ropke,  Z. Phys. A {\bf 351}, 295  (1995);  U. Lombardo, P. Nozieres,
P. Schuck, H.-J. Schulze,  and A. Sedrakian,  \prc {\bf 64},
064314 (2001); E. Babaev, \prb {\bf 63},  184514 (2001); B. Kerbikov,   Phys. Atom. Nucl. {\bf 65}, 1918 (2002); P. Castorina, G. Nardulli, and D. Zappala,  \prd {\bf 72}, 076006  (2005);  A. H. Rezaeian and H. J. Pirner, Nucl. Phys. A {\bf 779}, 197 (2006); J. Deng, A. Schmitt, and Q. Wang, Phys. Rev. D {\bf 76}, 034013 (2007);  T. Brauner, Phys. Rev. D {\bf 77}, 096006 (2008); M. Matsuzaki, 	Phys. Rev. D {\bf 82}, 016005 (2010); H. Abuki, G. Baym, T. Hatsuda, and N. Yamamoto, \prd {\bf 81} 125010 (2010).

\bibitem{NiAb05} Y. Nishida and H. Abuki, \prd {\bf 72}, 096004 (2005).

\bibitem{SeCl06} A. Sedrakian and
J. W. Clark,  chapter contributed to ``Pairing in Fermionic
Systems: Basic Concepts and Modern Applications", World
Scientific, nucl-th/0607028  (2006);  A. Sedrakian and J. W. Clark,  \prc {\bf 73}, 035803  (2006).

\bibitem{Ba04} E. Babaev,   \prd {\bf 70}, 043001 (2004).

\bibitem{Gl00} N. K. Glendenning,
Compact Stars, Nuclear Physics, Particle Physics and General
Relativity, Springer, New York (2000).

\bibitem{BaBa03} S. Banik and D. Bandyopadhyay, \prd {\bf 67},
123003 (2003).

\bibitem{Ban04} S. Banik, M. Hanauske, D. Bandyopadhyay,  and
W. Greiner, \prd {\bf 70}, 123004 (2004).

\bibitem{Ka04} J. I. Kapusta,   \prl {\bf 93}, 251801 (2004).

\bibitem{Ab06} H. Abuki,  Nucl. Phys. A {\bf 791}, 117 (2007).

\bibitem{SeSu91} E. Seidel and W.-M. Suen,  \prl {\bf 66}, 1659 (1991).

\bibitem{Al04} M. Alcubierre, J. A. Gonz\'alez,  and
M. Salgado,  \prd {\bf 70}, 064016 (2004).

\bibitem{Al04-1}   E. Seidel and W.-M. Suen, \prd, 42, 384 (1990);
  E. Seidel and W.-M. Suen,  \prl {\bf 72}, 2516 (1994); J. Balakrishna, E. Seidel,  and W.-M. Suen, \prd {\bf 58}, 104004 (1998);  F. S. Schunk and
E. W. Mielke,   Class. Quantum Grav. {\bf 20}, R301 (2003);   Y. F. Yuan, R. Narayan,  and
M. J. Rees,  Astrophys. J. {\bf 606}, 1112  (2004);  F. S. Guzman  and
L. A. Urena-Lopez, Astrophys. J. {\bf 645}, 814 (2006).

\bibitem{bosonstars} D.J. Kaup, Phys. Rev. {\bf  172}, 1331 (1968); R. Ruffini, S. Bonazzola, Phys. Rev. {\bf  187}, 1767 (1969); W. Thirring, Phys. Lett. B {\bf  127}, 27 (1983); J.D. Breit, S. Gupta, A. Zaks, Phys. Lett. B {\bf  140}, 329 (1984);  E.W. Mielke, F.E. Schunck, Nuc. Phys. B {\bf  564}, 185 (2000);  F.E. Schunck, E.W. Mielke, Class. Quantum Grav.  {\bf  20}, R301 (2003).

\bibitem{colpi} M. Colpi, S.L. Shapiro, I. Wasserman, Phys. Rev. Lett. {\bf 57}, 2485 (1986).

\bibitem{scalarfield} M.R. Baldeschi, G.B. Gelmini, R. Ruffini, Phys. Lett. B {\bf  122}, 221 (1983); S.J. Sin, Phys. Rev. D {\bf  50}, 3650 (1994); J. Lee, I. Koh, Phys. Rev. D {\bf  53}, 2236 (1996); F.E. Schunck, [astro-ph/9802258]; T. Matos, F.S. Guzm\'an, F. Astron. Nachr. {\bf 320}, 97 (1999);  P.J.E. Peebles, Astrophys. J. {\bf 534}, L127 (2000); W. Hu, R. Barkana, A. Gruzinov,
Phys. Rev. Lett. {\bf 85}, 1158 (2000); A. Arbey, J. Lesgourgues,
P. Salati, Phys. Rev. D {\bf 64}, 123528 (2001); M.P. Silverman,
R.L. Mallett, Gen. Rel. Grav. {\bf 34}, 633 (2002).

\bibitem{goodman} J. Goodman, New Astronomy {\bf 5}, 103 (2000).

\bibitem{arbey} A. Arbey, J. Lesgourgues, P. Salati, Phys. Rev. D {\bf 68}, 023511 (2003).

\bibitem{dark} C. G. B\"ohmer and T. Harko, JCAP \textbf{06}, 025 (2007);
 T. Fukuyama, M. Morikawa, and T. Tatekawa, JCAP {\bf 06}, 033 (2008); D. Boyanovsky, H. J. de Vega, and N. G. Sanchez, Phys. Rev. D \textbf{77}, 043518 (2008); J.-W. Lee, Phys. Lett. B {\bf 681}, 118 (2009);
 T. Fukuyama and M. Morikawa, Phys. Rev. D {\bf 80}, 063520 (2009);  T. Harko, MNRAS {\bf 413}, 3095 (2011); P.-H. Chavanis, arXiv1103.2698 (2011); P.-H. Chavanis, arXiv:1103.3219 (2011); T. Harko, JCAP {\bf 1105}, 022 (2011); T. Harko, Phys. Rev. D {\bf 83}, 123515 (2011); T. Rindler-Daller and P. R. Shapiro, arXiv:1106.1256 (2011).

\bibitem{Wa01}  X. Z. Wang,   \prd {\bf 64}, 124009 (2001).

\bibitem{Gr06} J. A. Grifols,  Astropart. Phys. {\bf 25}, 98 (2006).

\bibitem{Chav1} P.-H. Chavanis, arXiv:1103.2050 (2011).

\bibitem{Chav2} P.-H. Chavanis and L. Delfini, arXiv:1103.2054 (2011).

\bibitem{Ko00}  E. B. Kolomeisky, T. J. Newman, J. P.
Straley, and X. Qi,   \prl {\bf 85}, 1146 (2000).


\bibitem{Ch57}  S. Chandrasekhar,   An
introduction to the study of stellar structure, New York, Dover Publications (1957).

\bibitem{Ch33}  S. Chandrasekhar, MNRAS {\bf 93}, 390 (1933).

\bibitem{Ch92} H. F. Chau, K. S. Cheng,  and K. Y. Ding,  Astrophys. J. {\bf 399}, 217 (1992).


\bibitem{tooper} R.F. Tooper,   Astrophys. J. {\bf 140}, 434 (1964).

\bibitem{ijmpb} P.H. Chavanis, Int. J. Mod. Phys. B {\bf 20}, 3113 (2006).

\bibitem{OV} J. Oppenheimer and G. M. Volkoff, Phys. Rev. {\bf 55}, 374 (1939).

\bibitem{misner} C.W. Misner and H.S. Zapolsky, Phys. Rev. Lett. {\bf 12}, 635 (1964).

\bibitem{chavgen} P.H. Chavanis, Astron. Astrophys. {\bf 381}, 709 (2002).

\bibitem{aaiso} P.H. Chavanis, Astron. Astrophys. {\bf 381}, 340 (2002).

\bibitem{mt} D.W. Meltzer and K.S. Thorne, Astrophys. J. {\bf 145}, 514 (1966).

\bibitem{weinberg} S. Weinberg, Gravitation and Cosmology (John Wiley, 2002).

\bibitem{chavbh} P.H. Chavanis, Astron. Astrophys. {\bf 483}, 673 (2008).

\bibitem{st} S.L. Shapiro and S.A. Teukolsky, Black holes, white dwarfs and neutron stars (Wiley, New York, 1983).

\bibitem{aaantonov}  P.H. Chavanis, Astron. Astrophys. {\bf 451}, 109 (2006).

\bibitem{RR} C. E. Rhoades, Jr. and R. Ruffini, Phys. Rev. Lett. {\bf 32}, 324 (1974).

\bibitem{Kal} V. Kalogera and G. Baym, Astrophys. J. {\bf 470}, L61, (1996).

\bibitem{Bu}  H. A. Buchdahl,  Phys. Rev. {\bf 116}, 1027 (1959).

\bibitem{Lat} J. M. Lattimer and M. Prakash, to appear in Gerry Brown's Festschrift, Editor: Sabine Lee (World Scientific),	arXiv:1012.3208 (2010).

\bibitem{Dem} P. B. Demorest, T. Pennucci, S. M. Ransom, M. S. E. Roberts and J. W. T. Hessels,
Nature {\bf 467}, 1081 (2010).

\bibitem{black} O. Barziv, L. Karper, M. H. van Kerkwijk, J. H. Telging, and J. van Paradijs, Astron.
 Astrophys.{\bf  377}, 925 (2001);  H. Quaintrell, A. J. Norton, T. D. C. Ash, P. Roche, B. Willems, T. R. Bedding, I. K. Baldry, and R. P. Fender, Astron. Astrophys. {\bf 401}, 303 (2003); M. H. van Kerkwijk, R. Breton, and S. R. Kulkarni, 	arXiv:1009.5427 (2010).

\bibitem{ChHa03} K. S. Cheng and T. Harko, Astrophys. J. {\bf 596}, 451 (2003).

\bibitem{HaCh05} T. Harko and K. S. Cheng,   Astrophys. J. {\bf 622}, 1033 (2005).

\end{thebibliography}
\end{document}